\documentclass[lettersize,journal]{IEEEtran}
\usepackage{amsmath,amsfonts}
\usepackage{algorithmic}
\usepackage{algorithm}
\usepackage{array}
\usepackage{textcomp}
\usepackage{stfloats}
\usepackage{url}
\usepackage{verbatim}
\usepackage{graphicx}
\usepackage{subfigure}
\usepackage{float}
\usepackage{orcidlink}
\usepackage{latexsym}
\def\BibTeX{{\rm B\kern-.05em{\sc i\kern-.025em b}\kern-.08em
    T\kern-.1667em\lower.7ex\hbox{E}\kern-.125emX}}
\usepackage{balance}
\hypersetup{hidelinks,	
	colorlinks=true,	
	allcolors=black,	
	pdfstartview=Fit,	
	breaklinks=true}
\begin{document}
	
\title{Task Offloading for Vehicular Edge Computing Based on Improved Hotstuff under Parking Assistance}
\author{
	Guoling Liang \orcidlink{0009-0005-6694-720X},
	Chunhai Li \orcidlink{0000-0003-2543-1678}, ~\IEEEmembership{Member, ~IEEE},
	Feng Zhao \orcidlink{0000-0002-5730-2208}, ~\IEEEmembership{Member, ~IEEE},
	Chuan Zhang \orcidlink{0000-0001-7684-8540}, ~\IEEEmembership{Member, ~IEEE},
	Liehuang Zhu \orcidlink{0000-0003-3277-3887},  ~\IEEEmembership{Senior Member, ~IEEE}

	\thanks{This work was supported by the National Natural Science Foundation of China under Grant 62362013, the Guangxi Natural Science Foundation under Grant 2023GXNSFAA026294, and the National Natural Science Foundation of China under Grant 62232002  and Grant U21A20463.\textit{(Corresponding author: Chunhai Li.)}} 
	
	\thanks{Guoling Liang is with the Guangxi Engineering Research Center of Industrial Internet Security and Blockchain, Guilin University of Electronic Technology, Guilin 541004, China, and also with the School of Physics and Telecommunication Engineering, Yulin Normal University, Yulin 537000, China (e-mail: guolliang@mails.guet.edu.cn).}
	\thanks{Chunhai Li and Feng Zhao are with the Guangxi Engineering Research Center of Industrial Internet Security and Blockchain, Guilin University of Electronic Technology, Guilin 541004, China (e-mail: chunhaili@guet.edu.cn; zhaofeng@guet.edu.cn).}
	
	\thanks{Chuan Zhang and Liehuang Zhu are with the School of Cyberspace Science and Technology, Beijing Institute of Technology, Beijing 100081, China (e-mail: chuanz@bit.edu.cn; liehuangz@bit.edu.cn).}
}

\markboth{Journal of \LaTeX\ Class Files,~Vol.~18, No.~9, September~2020}%
{How to Use the IEEEtran \LaTeX \ Templates}

\IEEEpubid{\begin{minipage}{\textwidth}\ \centering
		Copyright \copyright 2024 IEEE. Personal use of this material is permitted. \\
		However, permission to use this material for any other purposes must be obtained 
		from the IEEE by sending a request to pubs-permissions@ieee.org.
\end{minipage}}

\maketitle

\begin{abstract}
Parked-assisted vehicular edge computing (PVEC) fully leverages communication and computing resources of parking vehicles, thereby significantly alleviating the pressure on edge servers. However, resource sharing and trading for vehicular task offloading in the PVEC environment usually occur between untrustworthy entities, which compromises the security of data sharing and transactions by vehicles and edge devices. To address these concerns, blockchain is introduced to provide a secure and trustworthy environment for offloading and transactions in PVEC. Nevertheless, due to the mobility of the vehicles, the processes of computing offloading and blockchain transactions are interrupted, which greatly reduces the reliability of the blockchain in edge computing process. In this paper, we propose a blockchain-based PVEC (BPVEC) offloading framework to enhance the security and reliability of the task offloading and transaction. Specifically, a consensus node selection algorithm based on the connected dominating set (CDS) is designed to improve the Hotstuff consensus according to parking time, computing capability and communication quality, which enhances blockchain reliability in computing offloading and transactions. Meanwhile, a Stackelberg game model, establishing the roadside units (RSUs) and parking vehicles (PVs) as leaders and the requesting vehicles (RVs) as follower, is utilized to optimize the offloading strategy and pricing. Subsequently, a BPVEC offloading strategy algorithm with gradient descent method is designed to maximize system revenue. Simulation results show that the proposed BPVEC offloading scheme is secure and reliable while ensuring maximum benefits.
\end{abstract}


\section{Introduction}
\IEEEPARstart{W}{ith} the computation-intensive and delay-sensitive applications developed, VEC servers are struggling with the limitation of their computing resources \cite{GaoKuang-606}. To counteract the challenge, a novel computing paradigm, Parked-assisted vehicular edge computing \cite{huang-8522034,zeng-000996147400001}, has emerged, utilizing the rich idle parking vehicles' resources to offer communication and computing services \cite{WOS:000719386001001,WOS:000696216800028}. According to investigation, a majority of vehicles spend most of their time parking in lots, garages, or driveways \cite{WOS:000531086200005}, which provides the conditions necessary for establishing relatively stable wireless communication among them and forming Vehicular Ad-Hoc Networks (VANETs), greatly expanding the network coverage of vehicle network. A PVEC framework allows users to offload computing tasks to parking vehicles through VANETs, thereby achieving efficient services and improving the service quality of vehicle applications. Furthermore, the PVEC task offloading aims to execute the tasks rapidly, conserve the device's battery power, enhance system performance and energy efficiency \cite{10594714,9615338}. To achieve this, PVEC task offloading optimizes performance by formulating an optimization model considering various factors, including energy consumption and latency. Therefore, user vehicles and parking vehicles, which often grapple with limitations of computing capabilities and battery power, can greatly benefit from PVEC task offloading. 

However, the task offloading and transaction process of PVEC may cause the leakage of sensitive data of honest entities \cite{WOS:000648799500001}, thereby posing privacy protection and transaction information security challenges. In addition, the process of collecting user data for training and prediction by PVEC's server will also introduce data security risks \cite{hu2023achieving, zhang2023achieving}. To address these challenges, researchers introduce blockchain technology that is well-known for its data security and integrity capabilities \cite{WOS:000929800100015}. With the open and tamper-proof consensus, blockchain ensures that all users can read the data in plaintext while preventing any unauthorized edits to the data \cite{zhang2023tdsc}. Additionally, the traceability and immutability of blockchain safeguard against malicious node attacks within the PVEC network, preserving the user security and privacy during task offloading \cite{LangTian-597}. Despite this, there remain some challenges to be addressed in integrating blockchain into the PVEC task offloading process, particularly interruptions of computing processes and blockchain \IEEEpubidadjcol transactions due to the mobility of vehicles, which affect the reliability of the blockchain during offloading. Moreover, ensuring system safety and reliability while maximizing benefits presents another challenge.

To address the aforementioned challenges, we propose a BPVEC offloading framework that guarantees the security and reliability of  task offloading and transaction. Additionally, a consensus node selection algorithm based on CDS is designed to improve Hotstuff consensus according to parking time, computing capability and communication quality. Subsequently, a Stackelberg game model with the RSU and the PV as leaders,  the RV as the follower is formulated to maximize system benefits. The primary contributions of this paper are summarized as follows:
\begin{enumerate}
	\item{We propose a new PVEC task offloading framework based on blockchain to  create a trustworthy environment for PVEC offloading. Within this framework, PVs and RSUs form a distributed network respectively to build a two-chain consortium blockchain system: subchain and main chain. The two chains save the pricing strategies and tasks of PV and RSU, respectively.} 
	\item{We design a CDS-Hotstuff consensus to enhance blockchain reliability in computing offloading and transaction.  Specifically, consensus nodes are selected from PVs based on their parking duration, computing capability, and communication quality, using a CDS-based node selection algorithm. Furthermore, the CDS-Hotstuff model is built to calculate the energy consumption required for consensus.}
	\item{We establish a Stackelberg game model with the RSUs and PVs as leaders and the RVs as the follower to optimize the offloading strategy. The CDS-Hotstuff consensus energy consumption is considered into the utility function to formulate related optimization problem. Additionally, a BPVEC offloading strategy algorithm based on the gradient descent method is developed to attain Nash equilibrium.}
	\item{We perform the simulation experiment, evaluating the system performance of proposed scheme and demonstrating that our scheme is secure and reliable, while ensuring the RV, PV and RSU benefits.} 
\end{enumerate}

The remaining sections of this paper are organized as follows. The following section reviews the related work on task offloading in PVEC and blockchain-based task offloading. Section III overviews the Hotstuff consensus mechanism and Section IV designs the BPVEC offloading system model. In Section V, we discuss the RV, PV, and RSU utility functions, and set up two-stage Stackelberg game. The Stackelberg game is analyzed, and the CDS-based consensus node selection algorithm and the BPVEC offloading strategy algorithm based on gradient descent method are designed in Section VI. Section VII evaluates the proposed scheme performance by experimental results. Finally, we make a conclusion for this paper in Section VIII.

\section{Related Work}
\subsection{Task Offloading in PVEC}
Task offloading of VEC seeks to optimize computing resources and advance resource utilization. However, the limitation of VEC server computing and communication resources greatly impacts the efficiency of task offloading. Therefore, HUANG et al. \cite{huang-8522034} first proposed PVEC, making the most of a large number of PV resources in parking lots or roadside to offer computing services for mobile vehicle. However, the authors only focused on the task price and economic benefit, when creating the utility function. To enhance this aspect, LI et al. \cite{WOS:000478957600024}  incorporated energy consumption into task offloading, following Huang's research, and suggested a new PVEC offloading scheme that reduced system costs. Nonetheless, neither scheme accounted for the impact of communication and computing delays on the efficiency of task offloading. Therefore, a novel task offloading strategy is designed to elevate the offloading performance in \cite{ma-9344808,fan-9709120}. To balance the impact of energy consumption and delay, SHEN et al. \cite{shen-22134959} introduced a  Mobile Edge Computing (MEC) offloading strategy leveraging roadside parking cooperation, created the resource sharing relationship among roadside vehicles, RSUs and cloud servers, and recast the cooperative offloading problem of computation tasks as a constrained optimization problem. By implementing the Hybrid Genetic Algorithm (HGA), they minimized both task delay and energy expenditure. Similarly, the studies in \cite{chen-7394689,10.1145/3514242} respectively put forward a parked vehicle-assisted MEC architecture and a resource management scheme, aiming to reduce delay and mobile device consumption, which improved task offloading efficiency, and reduced the burden on edge servers. However, the above works neglected the issue of resource pricing when considering energy consumption or latency. In subsequent research, the works in \cite{chen-10494795,qin-10637435} took into account the factors of latency, energy consumption, and resource pricing to establish task offloading, thereby decreasing offloading costs, delays, and energy utilization while improving task completion degree. Studies in 
\cite{zeng-000996147400001,pham-10373790} also contributed to similar work. On the other hand, in the discussion of task execution efficiency and resource utilization, \cite{wen-001137656300002} introduced a MEC offloading strategy based on collaborative roadside parking, significantly promoting the execution efficiency and energy utilization. Then a multi-stage Stackelberg game was developed to acquires the interactions between RVs, service providers (SPs), and PVs, with the goal of maximizing the welfare of all stakeholders in \cite{lei-10296002}. However, all the above works are about centralized task offloading. In order to construct a distributed task offloading architecture, the study in \cite{pham-9795919} tackled the complexity and privacy issues associated with centralized approaches, as well as the resource contention among Mobile Data Service Subscribers (MDSSs). It introduced a low-complexity distributed offloading scheme, which significantly increased the system's utilization rate. Nevertheless , actual task offloading is often dynamic, necessitating a more adaptable scheme that considers multiple factors, as \cite{XUE-2023103069} discussed. Consequently, incorporating reinforcement learning into PVEC task offloading, as suggested in \cite{ma-9964433,wang-10183330,xue-001248642900012}, improves the efficiency of dynamic task offloading. Although the above works have considered various factors to improve offloading efficiency and system performance, there is a lack of research on the security of offloading environment.

\subsection{Blockchain for Task Offloading}
As an emerging technology for secure sharing and synchronization, blockchain, by integrating with VEC, constructs a trustworthy distributed network that provides a secure environment for offloading and transactions by RVs. Accordingly, to ensure the security of offloading within VEC and to motivate RSUs and vehicles to share resources, the study in \cite{8923295} created a competitive model between RSUs and vehicles for data sharing and mining tasks. However, the reliance on POW consensus in this model greatly reduced task offloading efficiency and system performance. To address these issues, the work in \cite{ZhengLi-589} investigated the security and offloading problems within a cloud blockchain-based multi-vehicle Energy Consumption and Control Optimization (ECCO) system, aiming to enhance system performance while maintaining security. Similar studies are reported in \cite{10070689} and \cite{XiaoChen-595}. The above researches only focused on the improvement of the VEC task offloading system overall performance, neglecting the blockchain system performance. Therefore, several studies have directed their focus towards optimizing blockchain parameters to enhance offloading efficiency and system performance. A blockchain-based framework for delay-sensitive resource optimization, introduced in \cite{9120494},  is designed to optimize block size, generation time, offloading strategy, and computation frequency, thereby achieving computational offloading within delay constraints. In comparison, with the goal of minimizing service processing latency, reducing energy consumption and maximizing blockchain throughput, the offloading and migration optimization model was established in \cite{RenChen-590} to facilitate the inter-region trust and resource matching. YE et al. and QI et al. \cite{9500307,Qi-593} assessed the impact of consensus mechanism energy consumption and latency on computing offloading and optimized the blockchain, which reduced the comprehensive costs. In \cite{LangTian-597}, the author explored the secure handover of blockchain technology in cooperative computing offloading and proposed a cooperative computing offloading decision optimization method to optimize the latency of vehicular computation tasks. Nevertheless, the consensus algorithm is the major factor affecting the blockchain performance. As a result, enhancing blockchain performance can also involve development or improvement of consensus algorithm. Studies in \cite{HassijaChamola-601,LangTian-596,ShiDu-598} respectively improved on DAG and PBFT to optimize the blockchain to strengthen the VEC task offloading efficiency and system performance. Moreover, the integration of smart contracts in VEC can also create a trusted task offloading environment. A multi-step smart contract is proposed in \cite{WangYe-591} to realize secure resource sharing, preventing malicious behaviors of service requester and selfish vehicles. They also developed an incentive scheme based on smart contracts to motivate vehicles to contribute computing resources. The work in \cite{FuWang-599} presented a new blockchain-based offloading framework for Vehicle to Vehicle (V2V) and Vehicle to RSU (V2R) transactions, which ensured the secure and scalable data exchange and bandwidth transaction process, improved the quality of service for vehicles, and realized decentralized payments. On the other hand, HUANG et al. \cite{HuangYe-602} utilized a Stackelberg game to formulate and solve the smart contract design optimization problem, with the objective of minimizing user costs and improving offloading efficiency. However, the above works do not fully consider the impact of vehicle mobility, computing capability, and communication quality on blockchain during task offloading.

In summary, existing works on PVEC task offloading neglect the security aspects of offloading and transaction processes, as well as the reliability of blockchain within these processes. To address these issues, we introduce blockchain to design the PVEC task offloading framework and enhance the security of computation offloading and transaction. Moreover, a CDS-based consensus node selection strategy algorithm is presented to ensure the consensus reliability during the offloading process.

\section{Hotstuff Consensus Mechanism}
Blockchain, providing decentralized, secure, and reliable environment for PVEC task offloading and transactions, is primarily categorized into public chain and permissioned chain, i.e., consortium chain \cite{8985250}. Due to the advantages of low transaction cost, fast transaction execution speed, and excellent privacy protection in consortium chain \cite{WOS:001317694500083}, we select consortium blockchain to achieve our BPVEC offloading framework and aim to enhance its reliability in computing offloading and transaction by improving Hotstuff consensus. Therefore, this section provides an overview of Hotstuff.

\begin{figure}[h]
	\centering
	\includegraphics[scale=0.37]{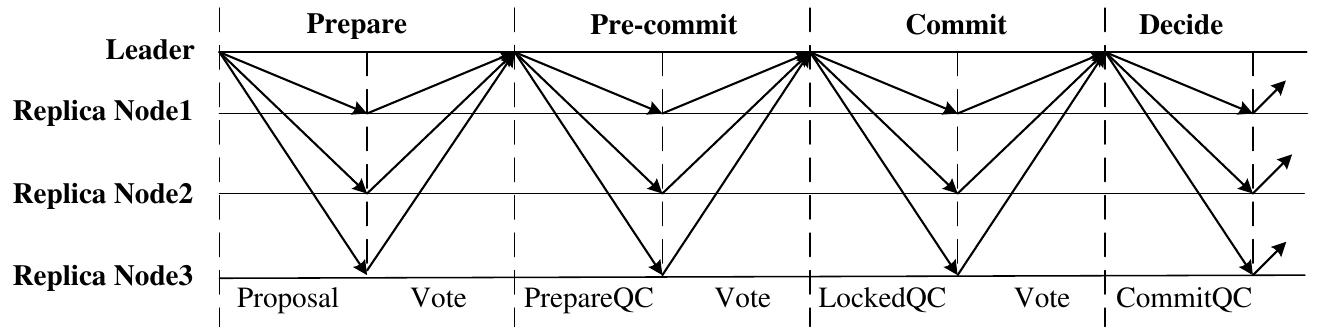}
	\caption{The Hotstuff consensus process}
	\label{fig1:figHotstuff}
\end{figure}

Hotstuff, an improvement on PBFT, is a Byzantine Fault Tolerance (BFT) consensus algorithm, that increases the authentication complexity to $O (N)$. It is divided into basic Hotstuff and chained Hotstuff. In this paper, we employ the basic Hotstuff; the process includes four phases: prepare , pre-commit, commit, and decision \cite{WOS:000570442000049}, as depicted in Fig.~\ref{fig1:figHotstuff}. Assuming $N$ is the number of consensus nodes and $F$ is the number of Byzantine nodes, then the consensus process unfolds as follows.

\textit{Prepare}: At the beginning of a new view, a new leader gathers \textit{new-view messages} from 2$F$+1 replicas, where the \textit{new-view message} carries the replica's highest view number, PrepareQC. Then, the leader selects the highest PrepareQC from these \textit{new-view messages} as HighQC and broadcasts the message $\langle\textit{prepare}, \textit{curProposal}, \textit{highQC}\rangle$ to replicas. After accepting the message of the current view leader, the replica nodes vote on the message and return the vote result message $\langle\textit{prepare}, \textit{m.node}, \bot\rangle$ and signature to the leader.

\textit{Pre-commit}: Since receiving the signatures from \textit{prepare message} of 2$F$+1 replica nodes, the leader creates PrepareQC and broadcasts $\langle\textit{pre-commit}, \bot, \textit{prepareQC}\rangle$. Then the replica nodes verify the message and send the \textit{pre-commit messages} $\langle\textit{pre-commit}, \textit{m.justify .node}, \bot\rangle$ and signature to the leader.

\textit{Commit}: Similar to the above phase, the leader aggregates 2$F$+1 the \textit{pre-commit messages} of replicas to create Pre-commitQC, then broadcasts $\langle\textit{commit}, \bot, \textit{pre-commitQC}\rangle$ to the replica nodes for verification. The replica nodes verify this and sign, refresh their own LockQC and return $\langle\textit{commit}, \textit{m.justify .node}, \bot\rangle$ to the leader. 

\textit{Decide}: Upon obtaining 2$F$ + 1 \textit{commit messages}, the leader forms CommitQC and broadcasts the \textit{decision message} $\langle\textit{decide}, \bot, \textit{commitQC}\rangle$. The replica nodes execute the block transactions, increase the \textit{viewNumber} and start a new view.
\section{System Model}
In this section, the BPVEC offloading system framework is described in the first three sections. then the improvement Hotstuff consensus scheme for the BPVEC task offloading is demonstrated in the last section. For reference, the main notations are summarized in Table~\ref{tab:Notation}.
\begin{table*}[h]
	\centering
	\caption{Explanation of notations}
	\label{tab:Notation}
	\begin{tabular}{|l|l|}
		\hline
		Notation & Explanation\\
		\hline
		$\mathcal{B}_{p}$ & PV consensus nodes set\\
		\hline
		$C^{exe}_{pk}, C^{exe}_{rj}$ & Number of CPU cycles per unit task size required for RSU and PV computing\\
		\hline
		$d_{0}$ & Wireless far-field reference distance\\
		\hline
		$d_{i,j}, d_{i,k}$ & Distance from RV to RSU or PV\\
		\hline
		$D_{qi}$ & Task size for RV\\
		\hline
		$D_{Bv}$ & Block size\\
		\hline
		$E^{BC}_{pa}, E^{BC}_{RSU}$ & PV and RSU consensus energy consumption\\
		\hline
		$f_{pk} , f_{rj}$ & Computational capacity of PV and RSU\\
		\hline
		$N_{0}$ & Noise power of the wireless channel\\
		\hline
		$P_{t}$ & Wireless transmission power\\
		\hline
		$p_{pai} , p_{RSUi}$ & PV and RSU price per unit task size\\
		\hline
		$p^{stay}_{k}$ & Vehicle $k$ remaining parking time\\
		\hline
		$Q_{k}$ & Node quality factor for PV\\
		\hline
		$R_{i,j} , R_{i,k}$ &	Transmission rate between RV and RSU or PV\\
		\hline
		$T_{maxi}$ &	Maximum tolerance time for RV\\
		\hline
		$T_{pai} , T_{RSUi}$ &	Time spent PV and RSU processing computational tasks\\
		\hline
		$W_{b}$ &	Wireless communication bandwidth\\
		\hline
		$\alpha$ &	RV satisfaction factor\\
		\hline
		$\beta$ &	The requirement of CPU cycles for generating or verifying signatures\\
		\hline
		$\delta$ &	Path loss factor\\
		\hline
		$\varepsilon_{RSUi}$ &	Proportion of task offloading to RSU by RV\\
		\hline
		$\eta$ &	Transceiver decision factor\\
		\hline
		$\theta$ &	The requirement of CPU cycles for generating or verifying the Message Authentication Code (MAC)\\
		\hline
		$\kappa_{v} , \kappa_{r}$ &	Capacitive switching coefficients for vehicles and RSUs\\
		\hline
		$\xi_{v} ,\xi_{r}$ &	Energy consumption per unit task size for vehicles and RSUs\\
		\hline
		$\phi_{pk} , \phi_{rj}$ &	PV and RSU computing capacity proportion\\
		\hline
		$\varpi$ & Average transaction size of blockchain\\
		\hline
	\end{tabular}
\end{table*}
\subsection{BPVEC Offloading System Framework}
The BPVEC offloading system framework is depicted as Fig.~\ref{fig2:system framework}. The system consists of RVs, PVs, RSUs, and VEC servers that are connected to RSUs. RSUs and PVs constitute blockchain networks respectively to realize distributed computational offloading and transaction services. The RSUs build a main chain network for recording global transactions. The PVs maintain a sunchain, and after completing consensus, they upload the blocks to the RSUs while retaining only the block header information. When an RV requires computing service, it sends a request message to an RSU or PV. Once RSU or PV verifies the legitimacy of the RV, the RV determines the offloading decision according to the pricing of RSU and PV. Following this, the RSU and PV carry out the transaction of computing service according to the offloading decision, then package and upload the record of the computation offloading and transactions to the blockchain system through consensus mechanism. This paper focuses on RSU and PV pricing strategy, RV’s offloading strategy, and how to combine with the consensus mechanism to realize distributed computing offloading.

\subsection{Network Model}
\begin{figure}[h]
	\centering
	\includegraphics[scale=0.43]{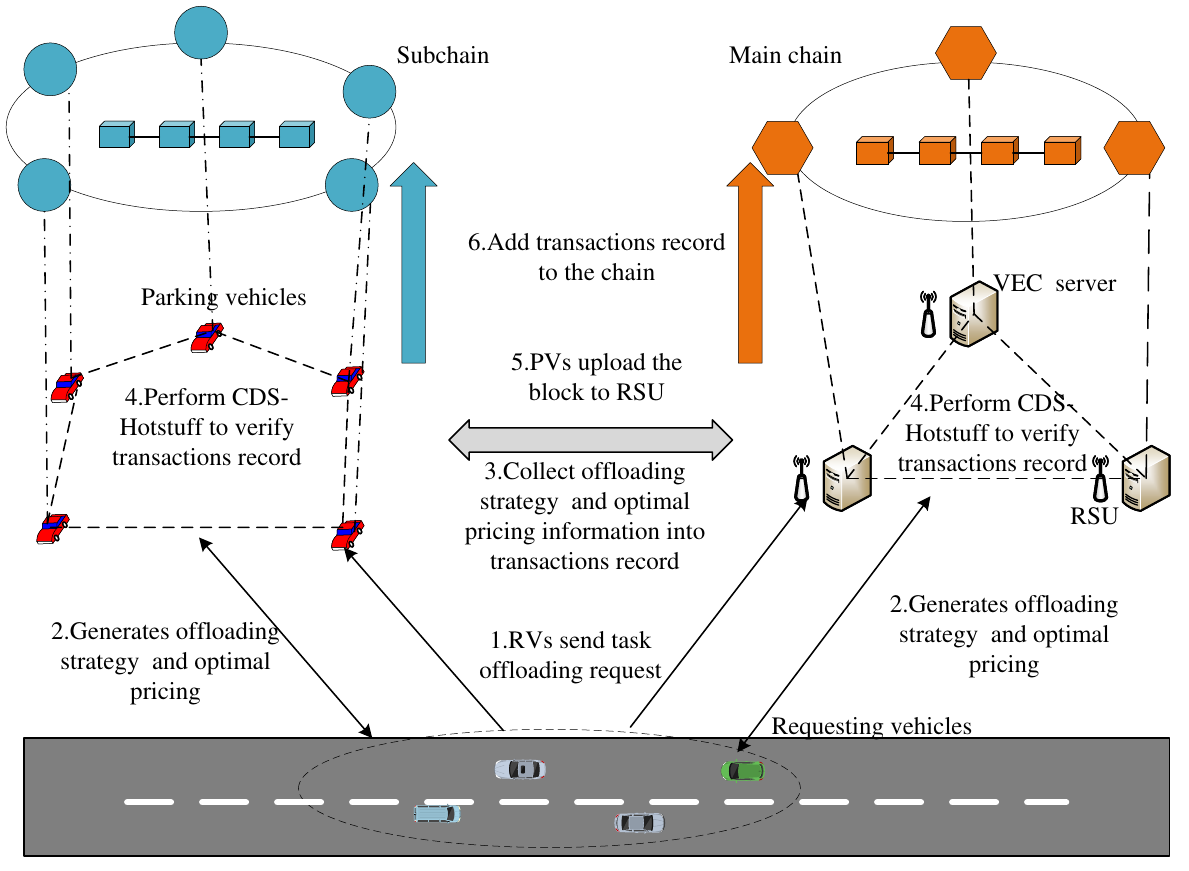}
	\caption{BPVEC offloading system framework}
	\label{fig2:system framework}
\end{figure}

As depicted in Fig.~\ref{fig2:system framework}, the region encompasses \textit{V} RVs, \textit{R} RSUs and \textit{P} PVs. The set of RVs is denoted by $\mathcal{V}$=$\{$1, 2, ... , \textit{V}$\}$, the set of RSUs by $\mathcal{R}$=$\{$1, 2, ... , \textit{R}$\}$ , and the set of PVs by $\mathcal{P}$=$\{$1, 2, ... , \textit{P}$\}$. According to \cite{MaZhao-611,MaTrivedi-612}, the transmission rates between RVs and RSUs, PVs are represented as
\begin{equation}
	\begin{array}{l}
		R_{i,j}  = W_b  \cdot \log \left( {1 + \frac{{P_t \eta (\frac{{d_0 }}{{d_{i,j} }})^\delta  }}{{N_0 }}} \right) \\ 
		R_{i,k}  = W_b  \cdot \log \left( {1 + \frac{{P_t \eta (\frac{{d_0 }}{{d_{i,k} }})^\delta  }}{{N_0 }}} \right) \\ 
	\end{array}
\end{equation}
Where $i\in\mathcal{V}$, $j\in\mathcal{R}$, $k\in\mathcal{P}$, $W_{b}$ signifies the wireless communication bandwidth, $P_{t}$ denotes the transmission power, $\eta$ is the transceiver decision coefficient. $d_{0}$ and $d_{i,j}, d_{i,k}$ represent the far-field reference distances, the distances from RV to the RSU or PV, respectively. $N_{0}$ is the noise power, and $\delta$ is the path loss factor.

\subsection{Parking Model}
The departure of PVs leads to the interruption of the task computation and consensus, hence, PVs with longer parking time should be selected as computation and consensus nodes. According to \cite{ReisSargento-619}, the cumulative probability density function of PV’s estimated residence time is
\begin{equation}
	\begin{array}{l}
		F(t_p ,t_a ) = \frac{{D_{1,t_a } \gamma (\kappa _{s,t_a } ,t_p /\theta _{s,t_a }^{\kappa _{s,t_a } } )}}{{\Gamma (\kappa _{s,t_a } )}} \\ 
		\begin{array}{*{20}c}
			{} & {}  \\
		\end{array} + \frac{{D_{2,t_a } \gamma (\kappa _{l,t_a } ,t_p /\theta _{l,t_a }^{\kappa _{l,t_a } } )}}{{\Gamma (\kappa _{l,t_a } )}} \\ 
	\end{array}
\end{equation}

In equation (2), $t_{p}>0$ and $t_{a}$$\in$$\{$0, 1, 2, ... , 23$\}$. $\kappa_{s,t_{a}}, \theta^{\kappa_{s,t_{a}}}_{s,t_{a}}$ represent the shape and scale parameters of the first Gamma distribution, characterizing the short-term parking behavior of PVs. $\kappa_{l,t_{a}}, \theta^{\kappa_{l,t_{a}}}_{l,t_{a}}$ are the parameters of the second Gamma distribution, describing the long-term parking behavior. $D_{1,t_{a}}$ and $D_{2,t_{a}}$ are weighting coefficients of the short-term and long-term Gamma distribution respectively, and satisfy $D_{1,t_{a}}+D_{2,t_{a}}=1$. In addition,  $\Gamma(\bullet)$ denotes the Gamma function and $\gamma(\bullet)$ is lower incomplete Gamma function. The parameters values are obtained through fitting the parking time data set, as detailed in the appendix of  \cite{ReisSargento-619}.

Let $t$ represent the current time slot and $t^{p}_{k}$ be the time that vehicle $k$ has parked. From equation (2), the probability that PV$k$ may park at least $\tau$ when it has parked $t^{p}_{k}$ at time $t$ is express as equation (3).

\begin{figure*}[hb] 
	\centering 
	\hrulefill 
	\vspace*{8pt} 
	\begin{equation}
		\begin{aligned}
			p^{stay}_{k}&=p[t>t^{p}_{k}+\tau|t>t^{p}_{k}]\\ 
			&=\left.\dfrac{D_{1,t_{a}}\gamma(\kappa_{s,t_{a}},\dfrac{t^{p}_{k}+\tau}{\theta^{\kappa_{s,t_{a}}}_{s,t_{a}}})\Gamma(\kappa_{s,t_{a}})+D_{2,t_{a}}\gamma(\kappa_{l,t_{a}},\dfrac{t^{p}_{k}+\tau}{\theta^{\kappa_{l,t_{a}}}_{l,t_{a}}})\Gamma(\kappa_{l,t_{a}})-\Gamma(\kappa_{s,t_{a}})\Gamma(\kappa_{l,t_{a}})}{D_{1,t_{a}}\gamma(\kappa_{s,t_{a}},\dfrac{t^{p}_{k}}{\theta^{\kappa_{s,t_{a}}}_{s,t_{a}}})\Gamma(\kappa_{s,t_{a}})+D_{2,t_{a}}\gamma(\kappa_{l,t_{a}},\dfrac{t^{p}_{k}}{\theta^{\kappa_{l,t_{a}}}_{l,t_{a}}})\Gamma(\kappa_{l,t_{a}})-\Gamma(\kappa_{s,t_{a}})\Gamma(\kappa_{l,t_{a}})}\right. 
		\end{aligned}
	\end{equation}
\end{figure*}

Let $p_{th}$ be the probability threshold of the parking duration. To enhance the computation and consensus reliability, the PVs with $p^{stay}_{k} \geq p_{th}$ should be selected to serve as the computation and consensus nodes.
\subsection{CDS-Hotstuff Model}
This section examines the consensus energy consumption, illustrated by the example of PVs executing the consensus mechanism. Since RSUs employ the same mechanism, their energy consumption model is similar to the PVs. Let $\beta$ and $\theta$ represent the CPU cycles for generating or verifying the signature and MAC, respectively, and let $B_{p}$ be the consensus nodes selected from PV nodes by CDS-based node selection algorithm, as described in Algorithm 2. If \textit{Bpl} is the leader node and \textit{Bpm} is the replica node, then their relationship is defined as $Bpl, Bpm\in\mathcal{B}_{p}=$ $\{$1, 2, 3, ... , $Bp$$\}$$\subseteq\mathcal{P}$. Furthermore, $F$ is the Byzantine nodes. Let $D_{Bv}$ and $\varpi$ denote the block size and the average transaction size, respectively. Drawing on the basic Hotstuff from Section III, the CDS-Hotstuff is divided into five phases.

\textit{New-view phase}: During this phase, the leader is tasked with verifying transactions and signatures from RVs, as well as the PrepareQC message and signature from the replica nodes. Since CDS-Hotstuff utilizes aggregate signature, which require the validation of $2F+1$ PrepareQC messages, the required number of CPU cycles is $O_{Bpl}^{req}  = \frac{{D_{Bv} (\beta  + \theta )}}{\varpi } + \beta  + (2F + 1)\theta $. In contrast, a replica node, responsible for generating a PrepareQC message and signature, requires $O_{Bpm}^{req}  = \beta  + \theta $ CPU cycles.

\textit{Prepare phase}: After verifying the transactions and signatures and confirming the new-view message, the leader generates a new block on the secure branch of HighQC, signing and sending it as a proposal to the replica nodes.  Therefore, the CPU cycles required by the leader in this phase are calculated by $O_{Bpl}^{prep}  = \beta  + \theta $
. A replica node, tasked with verifying the transactions and the block, casting a vote on the proposal, and signing and returning the vote message to the leader, requires $O_{Bpm}^{prep}  = 2\beta  + 2\theta  + \frac{{D_{Bv} }}{\varpi }(\beta  + \theta )$ CPU cycles.

\textit{Pre-commit phase}: Upon receiving votes from replica nodes regarding the proposal message, the leader aggregates the replicas' signatures into a new signature, creating a PrepareQC that is stored locally. Concurrently, the leader authenticates the votes and signatures from the replica nodes and broadcasts a pre-commit message to them. The leader's CPU cycles for this process are $O_{Bpl}^{pre\_com}  = 2\beta  + (2F + 1)\theta$. Meanwhile, validating the pre-commit message and casting a vote requires $O_{Bpm}^{pre\_com}  = 2\beta  + 2\theta $ CPU cycles for a replica node.

\textit{Commit phase}: Similar to the pre-commit phase, the leader receives votes from the replica nodes, validates them, and puts the Pre-commitQC into a commit message to broadcast to the replica nodes. Therefore, the CPU cycles for the leader and a replica node can be calculated by $O_{Bpl}^{com}  = 2\beta  + (2F + 1)\theta $ and $ O_{Bpm}^{com}  = 2\beta  + 2\theta$, respectively.

\textit{Decide phase}: Upon receiving 2$F$+1 commit-vote messages, the leader aggregates them into a CommitQC and broadcasts a decide message to the replica nodes. After accepting the CommitQC in the decide message, the replica node treats the current proposal as a definite message, and then executes the transaction on the branch that has been determined. At the same time, the view number is incremented to start a new phase. Therefore, the CPU cycles for the leader are $O_{Bpl}^{dec}  = 2\beta  + (2F + 1)\theta $, while the cycles for the replica nodes are negligible.

As analyzed above, the energy consumption for generating and verifying signatures and MACs in the consensus phase can be denoted as
\begin{equation}
	\begin{array}{l}
		E_{pa}^v  = \kappa _v f_{pl} ^2 \left( {O_{Bpl}^{req}  + O_{Bpl}^{pre}  + O_{Bpl}^{pre\_com}  + O_{Bpl}^{com}  + O_{Bpl}^{dec} } \right) \\ 
		+ \kappa _v f_{pm} ^2 \sum\limits_{\left\{ {Bpm \in {\cal B}_p \backslash Bpl} \right\}} {\left( {O_{Bpm}^{req}  + O_{Bpm}^{prep}  + O_{Bpm}^{pre\_com}  + O_{Bpm}^{com} } \right)}  \\ 
	\end{array}
\end{equation}

In addition, the transmission energy consumption in the consensus phase may be determined by formula (5).
\begin{equation}
	\begin{array}{l}
		E_{pa}^T  = \sum\limits_{\left\{ {Bpm \in {\cal B}_p \backslash Bpl} \right\}} {\frac{{4D_{B_v } }}{{R_{Bpm,Bpl} }}} \cdot P_t  \\ 
		\begin{array}{*{20}c}
			{} & {}  \\
		\end{array} + \sum\limits_{\left\{ {Bpm \in {\cal B}_p \backslash Bpl} \right\}} {\frac{{4D_{B_v } }}{{R_{Bpl,Bpm} }}} \cdot P_t  \\ 
	\end{array}
\end{equation}

Therefore, the total energy consumption due to the consensus mechanism for $V$ RVs should be calculated by
\begin{equation}
	E_{pa}^{{\rm{BC}}} {\rm{ = }}\frac{V}{{D_{Bv} /\varpi }}(E_{pa}^v  + E_{pa}^T )
\end{equation}

\section{Stackelberg Game Construction for Optimal Offloading Strategy}
\subsection{RV Utility Function}
A RV utility function can be defined as the difference between the RV's satisfaction with the computing service time and the completing computing task cost, i.e., 
\begin{equation}
	\begin{array}{l}
		U_{ci}  = F_i (T_i ) - H_{pai} (p_{pai} )\\ 
		\begin{array}{*{20}c}
			{} & {}  \\
		\end{array} - H_{RSUi} (p_{RSUi} ) - L_{ci} (\varepsilon _{RSUi} )\\
	\end{array}
\end{equation}

Here, $F_{i} (T_{i})$ signifies the RV satisfaction function regarding the computed service time. $H_{pai}(p_{pai})$ and $H_{RSUi}(p_{RSUi)}$ correspond to the pricing functions for the PV and RSU, respectively. $L_{ci} (\varepsilon_{RSUi})$ denotes the communication energy consumption function for task offloading by RV. $F_{i} (T_{i})$ is delineated by equation (8).
\begin{equation}
	F_i (T_i ) = \alpha (T_{\max i}  - T_i ^2 )	
\end{equation}

Where $\alpha$ is the satisfaction coefficient, $T_{maxi}$ is the maximum tolerance time of computing service for RV$i$, and $T_i  = \max (T_{pai} ,T_{RSUi} )$. $T_{pai}$ and $T_{RSUi}$ represent the total task completion times for the PV and RSU, respectively, as detailed by equation (9) and (10).
\begin{equation}
	T_{pai}  = \mathop {\max }\limits_k \left( {\frac{{(1 - \varepsilon _{RSUi} )D_{qi} \phi _{pk} C_{pk}^{exe} }}{{f_{pk} }} + \frac{{(1 - \varepsilon _{RSUi} )D_{qi} }}{{R_{i,k} }}} \right)
\end{equation}
\begin{equation}
	T_{RSUi}  = \mathop {\max }\limits_j \left( {\frac{{\varepsilon _{RSUi} D_{qi} \phi _{rj} C_{rj}^{exe} }}{{f_{rj} }} + \frac{{\varepsilon _{RSUi} D_{qi} }}{{R_{i,j} }}} \right)
\end{equation}

In equation (9) and (10), $\varepsilon _{RSUi}$ is defined as the proportion of the task size that RV$i$ offloads to the RSU, $D_{qi}$ is the task size that RV$i$ offloads, $\phi _{pk}, \phi _{rj}$ denote the proportion of the computational capacity of PV$k$ and RSU$j$ in all the nodes of PVs and RSUs, respectively. The CPU cycles required for computation per unit task size for PV and RSU are represented by $C_{pk}^{exe}$ and $C_{rj}^{exe}$, respectively. $f_{pk} , f_{rj}$ denote the computational capacity of PV$k$ and RSU$j$, respectively. $R_{i,k} , R_{i,j}$ are the transmission rate of RV$i$ offloading to PV$k$ and RSU$j$, respectively. If a RV has a computing task offloaded to the PV, the service fee for the PV computing service should be paid. The PV pricing function is given by
\begin{equation}
	H_{pai} (p_{pai} ) = (1 - \varepsilon _{RSUi} )D_{qi} p_{pai} 
\end{equation}

Here, $p_{pai}$ is the price per unit task size set by the PV to RV$i$. Similarly, the pricing function of the RSU is as equation (12).
\begin{equation}
	H_{RSUi} (p_{RSUi} ) = \varepsilon _{RSUi} D_{qi} p_{RSUi} 
\end{equation}

Where $p_{RSUi}$ is the price per unit task size set by the RSU to RV$i$. Moreover, $L_{ci}(\varepsilon _{RSUi})$ can be determined by (13).
\begin{equation}
	\begin{array}{l}
		L_{ci} (\varepsilon _{RSUi} ) = P_t \xi _v \frac{{\varepsilon _{RSUi} D_{qi} }}{{R_{i,j} }} \\ 
		\begin{array}{*{20}c}
			{} & {}  \\
		\end{array}\begin{array}{*{20}c}
			{} & {}  \\
		\end{array} + P_t \xi _v \frac{{(1 - \varepsilon _{RSUi} )D_{qi} }}{{R_{i,k} }} \\ 
	\end{array}
\end{equation}

Here, $ \xi _v $ represents the energy consumption per unit task size of vehicle processing. Substituting (11)-(13) into equation (7), the RV$i$ utility function is expressed as:
\begin{equation}
	\begin{array}{l}
		U_{ci}  = \alpha (T_{\max i}  - T_i ^2 ) \\ 
		\begin{array}{*{20}c}
			{} & {}  \\
		\end{array} - (1 - \varepsilon _{RSUi} )D_{qi} p_{pai}  - \varepsilon _{RSUi} D_{qi} p_{RSUi}  \\ 
		\begin{array}{*{20}c}
			{} & {}  \\
		\end{array} - \xi _v \left[ {P_t \frac{{\varepsilon _{RSUi} D_{qi} }}{{R_{i,j} }} + P_t \frac{{(1 - \varepsilon _{RSUi} )D_{qi} }}{{R_{i.k} }}} \right] \\ 
	\end{array}
\end{equation}

\subsection{PV Utility Function}
The PV utility function is composed of the total payment from RV to PV and the energy consumption generated by the PV task processing, as shown in equation (15).
\begin{equation}
	U_{pai}  = (1 - \varepsilon _{RSUi} )D_{qi} p_{pai}  - \xi _v E_{pai} 
\end{equation}

Where $\xi _v E_{pai}$ denotes the energy consumption cost of task processing. This energy consumption includes the consumption by computing and the consumption by consensus, i.e., $E_{pai}  = E_{pai}^{exe}  + E_{pai}^{BC}$, both of which are represented by the following two equations.
\begin{equation}
	E_{pai}^{exe}  = \sum\limits_{k = 1}^P {\kappa _v f_{pk}^2 (1 - \varepsilon _{RSUi} )D_{qi} \phi _{pk} C_{pk}^{exe} } 
\end{equation}
\begin{equation}
	E_{pai}^{{\rm{BC}}} {\rm{ = }}\frac{1}{{D_{Bv} /\varpi }}(E_{pa}^v  + E_{pa}^T )
\end{equation}

$\kappa _v $ is the vehicle capacitive switching factor. By equation (16) and (17), equation (15) can be expressed as
\begin{equation}
	\begin{array}{l}
		U_{pai}  = (1 - \varepsilon _{RSUi} )D_{qi} p_{pai}  \\ 
		\begin{array}{*{20}c}
			{} & {}  \\
		\end{array} - \sum\limits_{k = 1}^P {\xi _v \kappa _v f_{pk}^2 (1 - \varepsilon _{RSUi} )D_{qi} \phi _{pk} C_{pk}^{exe} }  \\ 
		\begin{array}{*{20}c}
			{} & {}  \\
		\end{array} + \frac{{\xi _v }}{{D_{Bv} /\varpi }}\left( {E_{pa}^v  + E_{pa}^T } \right) \\ 
	\end{array}
\end{equation}
\subsection{RSU Utility Function}
Similar to the PV utility function, the RSU utility function is expressed as
\begin{equation}
	U_{RSUi}  = \varepsilon _{RSUi} D_{qi} p_{RSUi}  - \xi _r E_{RSUi} 
\end{equation}

$\xi _r E_{RSUi}$ denotes the energy consumption cost incurred by processing tasks, while $\xi _r$ is the energy consumption per unit task size processed by the RSU, and $E_{RSUi}  = E_{RSUi}^{exe}  + E_{RSUi}^{BC}$, where
\begin{equation}
	E_{RSUi}^{exe} {\rm{ = }}\sum\limits_{j = 1}^R {\kappa _r f_{rj}^2 \varepsilon _{RSUi} D_{qi} \phi _{rj} C_{rj}^{exe} } 
\end{equation}
\begin{equation}
	E_{RSUi}^{{\rm{BC}}} {\rm{ = }}\frac{1}{{D_{Br} /\varpi }}(E_{RSU}^v  + E_{RSU}^T )
\end{equation}

$\kappa _r$ is the RSU capacitive switching coefficient. Therefore, equation (19) can be expressed as
\begin{equation}
	\begin{array}{l}
		U_{RSUi}  = \varepsilon _{RSUi} D_{qi} p_{RSUi}  \\ 
		\begin{array}{*{20}c}
			{} & {}  \\
		\end{array} - \sum\limits_{j = 1}^R {\xi _r \kappa _r f_{rj}^2 \phi _{rj} \varepsilon _{RSUi} D_{qi} C_{rj}^{exe} }  \\ 
		\begin{array}{*{20}c}
			{} & {}  \\
		\end{array} + \frac{{\xi _r }}{{D_{Br} /\varpi }}\left( {E_{RSU}^v  + E_{RSU}^T } \right) \\ 
	\end{array}
\end{equation}
\subsection{Utility Optimization Problem Form}
In this section, the utility functions of RV, PV and RSU are formulated according to the Stackelberg game. As outlined in Section IV, the task offloading process is outlined as follows: First, the RV sends its computing task requirements to the RSU and the PV when necessary. Then, the RSU and the PV learn each other’s strategies from the historical blockchain records and set prices based on this information and the task size. Finally, the RV determines the offloading strategy according to the prices of PV and RSU. Therefore, the above process conforms to the Stackelberg game with complete information.

As described above, the Stackelberg game model of BPVEC computing offloading can be split into two stages. At the first stage, PVs and RSUs as leaders determine their respective prices for computing services. At this stage, PVs and RSUs obtain information based on the historical records of the blockchain and compete with each other on price. Therefore, the pricing strategies are interdependent. At the second stage, the RV determines the offloading strategy based on the pricing of PVs and RSUs. The objective for an RV is to maximize its utility, which can be described by Problem 1.

Problem 1:
\begin{equation}
	\begin{array}{l}
		{\kern 1pt} {\kern 1pt} {\kern 1pt} {\kern 1pt} \mathop {\max }\limits_{\varepsilon _{RSUi} } U_{ci}  \\ 
		s.t.{\kern 1pt} {\kern 1pt} {\kern 1pt} {\kern 1pt} {\kern 1pt} {\kern 1pt} {\kern 1pt} {\kern 1pt} {\kern 1pt} {\kern 1pt} {\kern 1pt} T_{pai} ,T_{RSUi}  \le T_{\max i}  \\ 
		{\kern 1pt} {\kern 1pt} {\kern 1pt} {\kern 1pt} {\kern 1pt} {\kern 1pt} {\kern 1pt} {\kern 1pt} {\kern 1pt} {\kern 1pt} {\kern 1pt} {\kern 1pt} {\kern 1pt} {\kern 1pt} {\kern 1pt} {\kern 1pt} {\kern 1pt} {\kern 1pt} {\kern 1pt} {\kern 1pt} {\kern 1pt} 0 \le \varepsilon _{RSUi}  \le 1 \\ 
	\end{array}
\end{equation}

The PV and RSU also need to maximize their utility, which can be formulated as Problems 2 and 3.

Problem 2:
\begin{equation}
	\begin{array}{l}
		{\kern 1pt} {\kern 1pt} {\kern 1pt} \mathop {\max }\limits_{p_{pai} } U_{pai}  \\ 
		s.t.{\kern 1pt} {\kern 1pt} {\kern 1pt} {\kern 1pt} {\kern 1pt} {\kern 1pt} {\kern 1pt} {\kern 1pt} {\kern 1pt} {\kern 1pt} {\kern 1pt} U_{pai}  \ge 0 \\ 
	\end{array}
\end{equation}

Problem 3:
\begin{equation}
	\begin{array}{l}
		{\kern 1pt} {\kern 1pt} {\kern 1pt} {\kern 1pt} \mathop {\max }\limits_{p_{RSUi} } U_{RSUi}  \\ 
		s.t.{\kern 1pt} {\kern 1pt} {\kern 1pt} {\kern 1pt} {\kern 1pt} {\kern 1pt} {\kern 1pt} {\kern 1pt} {\kern 1pt} {\kern 1pt} {\kern 1pt} {\kern 1pt} {\kern 1pt} U_{RSUi}  \ge 0 \\ 
	\end{array}
\end{equation}
\section{BPVEC Task Offloading Strategy Scheme}
This section designs the CDS-based node selection algorithm to address the issues of PV duration and communication quality that may lead to interruption of the computation and consensus process. Subsequently, the Problem 1-3 are solved by using the backward induction. And finally, a BPVEC offloading strategy algorithm using the gradient descent method is proposed to realize the Stackelberg equilibrium.

\subsection{CDS-based Node Selection Algorithm}
When the parking duration of a PV is insufficient, it risks interrupting both computing and consensus processes. In this paper, the probability that PV$k$ can continue to park for a longer time than $\tau_{th}$ is calculated by equation (3). And nodes are selected as computing nodes based on the criterion $p_k^{stay}  \ge p_{th} $, resulting in the set $\mathcal{P}_C$ =$\{$1, 2, ..., $P_C$$\}$$\subseteq\mathcal{P}$. Subsequently, based on the CDS algorithm, the nodes with strong communication quality and computing ability are selected from $\mathcal{P}_C$ as consensus nodes to participate in Hotstuff consensus, thereby enhancing the consensus reliability. According to the network model described in Section IV, the signal-to-noise ratio between PV$k$ and PV$m$ is determined by equation (26).
\begin{algorithm}[]
	\caption{CDS-based node selection algorithm}
	\begin{algorithmic}[1]	
		\STATE \textbf{Input:} set of PV $\mathcal{P}$=$\{$1,2,...,$P$$\}$;
		\STATE \textbf{Output:} set of computing nodes for PV $\mathcal{P}_C$; set of consensus nodes for PV $\mathcal{B}_p$;
		\STATE \textbf{Initialization:} $\mathcal{P}_C=\emptyset$, $\mathcal{B}_p=\emptyset$;
		\STATE $p^{stay}_i\gets$equation (3)
		\IF {$p_i^{stay}  \ge p_{th}$}
		\STATE $\mathcal{P}_C=\mathcal{P}_C \cup i$
		\ENDIF
		\STATE $SNR_{i,j}\gets$equation (26)
		\IF {$SNR_{i,j } \ge SNR_{th}$}  
		\STATE $A_{i,j}=1;A_{j,i}=1$
		\ENDIF
		\STATE $Q_k\gets$equation (27)
		\FOR{$i=1:P_{C}$ }
		\STATE $index \gets max(Q)$
		\IF{covered$i$ == 0}
		\STATE $head=head \cup i$
		\ENDIF
		\FOR{$j=1:P_C$}
		\IF{$A_{index,j} ==1$}
		\STATE $covered_j=1$
		\ENDIF
		\ENDFOR		 
		\ENDFOR
		\FOR{$i=1:P_C$}
		\FOR{$j=i:P_C$}
		\FOR{$k=1:P_C$}
		\FOR{$n=1:P_C$}
		\IF{node$i,j$ is header and there is a common neighbor node$k$ or $i, j$ are connected by $k,n$} 
		\STATE $\mathcal{B}_p=\mathcal{B}_p\cup\{i,j,k,n\}$
		\ENDIF
		\ENDFOR
		\ENDFOR
		\ENDFOR
		\ENDFOR
		\label{code:CDS}
	\end{algorithmic}
\end{algorithm}
\begin{equation}
	SNR_{k,m}  = \frac{{P_t \eta (\frac{{d_0 }}{{d_{k,m} }})^\delta  }}{{N_0 }}
\end{equation}

Let $f_{pk}$ be the computing capacity of PV$k$ , then the node quality is calculated by equation (27). Where $w_1, w_2$ are the weight factors of communication quality and computing capacity, respectively. The node selection algorithm is shown as Algorithm 1.
\begin{equation}
	Q_k  = w_1  \cdot \frac{{\sum\limits_{m \in \mathcal{P}_C ,m \ne k} {SNR_{k,m} } }}{{\sum\limits_{k = 1}^{P_C } {\sum\limits_{m \in \mathcal{P}_C,m \ne k} {SNR_{k,m} } } }} + w_2  \cdot \frac{{f_{pk} }}{{\sum\limits_{k = 1}^{P_C } {f_{pk} } }}
\end{equation}

\subsection{Backward Induction for Stackelberg Equilibrium}
Problem 1-3 can be resolved in two stages: In the first stage, RSU and PV respectively price RV$i$'s task to maximize its utility function. The second stage determines the offloading strategy for the RV based on the pricing of RSU and PV to maximize the utility function. In this paper, we analyze the two stages by backward induction.

Stage 2: Analysis of offloading strategies for RV

With the pricing of RSUs and PVs determined, RVs need to decide the offloading task ratio $\varepsilon _{RSUi}$, and maximize the utility function. In the case of $T_i=T_{pai}$, suppose that the PV$k$ serves time is the maximum, and let $\Gamma _{pai}  = \mathop {\max }\limits_k \frac{{D_{qi} \left( {\phi _{pk} C_{pk}^{exe} R_{i,k}  + f_{pk} } \right)}}{{f_{pk} R_{i,k} }}$
, then according to equation (9), we have $T_{pai}  = (1 - \varepsilon _{RSUi} )\Gamma _{pai}$. The derivative of utility function can be calculated as
\begin{equation}
	\frac{{\partial U_{ci} }}{{\partial \varepsilon _{RSUi} }} = 2\alpha (1 - \varepsilon _{RSUi} )\Gamma _{pai}^2  + A
\end{equation}

Among them, $A = \left[ {D_{qi} p_{pai}  + \xi _v P_t \frac{{D_{qi} }}{{R_{i,k} }}} \right] - \left[ {D_{qi} p_{RSUi}  + \xi _v P_t \frac{{D_{qi} }}{{R_{i,j} }}} \right]$.

Therefore, $\frac{{\partial ^2 U_{ci} }}{{\partial \varepsilon _{RSUi}^2 }} =  - 2\alpha \Gamma _{pai}^2  < 0$.

When $T_i=T_{RSUi}$, set $\Gamma _{RSUi}  = \mathop {\max }\limits_j \frac{{D_{qi} \left( {\phi _{rj} C_{rj}^{exe} R_{i,j}  + f_{rj} } \right)}}{{f_{rj} R_{i,j} }}$, according to equation (10) we have $T_{RSUi}  = \varepsilon _{RSUi} \Gamma _{RSUi}$. The derivative of the utility function can be expressed as

\begin{equation}
	\frac{{\partial U_{ci} }}{{\partial \varepsilon _{RSUi} }} =  - 2\alpha \varepsilon _{RSUi} \Gamma _{RSUi}^2  + A
\end{equation}

Therefore, $\frac{{\partial ^2 U_{ci} }}{{\partial \varepsilon _{RSUi}^2 }} =  - 2\alpha \Gamma _{RSUi}^2  < 0$.

In summary, $U_{ci}$ is a strictly concave function, rendering the Problem 1 is a convex optimization problem. The optimal task offloading ratio $\varepsilon _{RSUi}$  is derived by using Karush-Kuhn-Tucker (KKT) conditions to solve Problem 1, as shown in equation (30).
\begin{equation}
	\varepsilon _{RSUi}^*  = \left\{ \begin{array}{l}
		1 + \frac{A}{{2\alpha \Gamma _{pai}^2 }}\begin{array}{*{20}c}
			{\begin{array}{*{20}c}
					{} & {}  \\
			\end{array}} & {T_i }  \\
		\end{array} = T_{pai}  \\ 
		\frac{A}{{2\alpha \Gamma _{RSUi}^2 }}\begin{array}{*{20}c}
			{\begin{array}{*{20}c}
					{} & {}  \\
			\end{array}} & {T_i }  \\
		\end{array} = T_{RSUi}  \\ 
		\frac{{\Gamma _{pai} }}{{\Gamma _{RSUi}  + \Gamma _{pai} }}\begin{array}{*{20}c}
			{} & {T_i  = T_{RSUi} }  \\
		\end{array} = T_{pai}  \\ 
	\end{array} \right.
\end{equation}

Stage 1: Pricing strategy analysis for RSU and PV

At this stage, the RSU and PV determine their pricing strategies and maximize their respective utility functions. Since the pricing decisions of the RSU and PV interact with each other, this stage can be regarded as a non-cooperative game between them, where they adjust their prices until achieving Nash equilibrium.

Theorem 1: A Nash equilibrium exists for the game between the RSU and PV.

The derivative of the PV and RSU utility functions is obtained by equation (31) and (32), respectively.
\begin{equation}
	\left\{ \begin{array}{l}
		\frac{{\partial U_{pai} }}{{\partial p_{pai} }} = (1 - \varepsilon _{RSUi}^* )D_{qi}  - D_{qi} p_{pai} \frac{{\partial \varepsilon _{RSUi}^* }}{{\partial p_{pai} }} \\ 
		\begin{array}{*{20}c}
			{} & {}  \\
		\end{array} + \xi _v \sum\limits_{k = 1}^P {\kappa _v f_{pk}^2 D_{qi} \phi _{pk} C_{pk}^{exe} \frac{{\partial \varepsilon _{RSUi}^* }}{{\partial p_{pai} }}}  \\ 
		\frac{{\partial ^2 U_{pai} }}{{\partial p_{pai}^2 }} =  - D_{qi} p_{pai} \frac{{\partial ^2 \varepsilon _{RSUi}^* }}{{\partial p_{pai}^2 }} - 2D_{qi} \frac{{\partial \varepsilon _{RSUi}^* }}{{\partial p_{pai} }} \\ 
		\begin{array}{*{20}c}
			{} & {}  \\
		\end{array} + \xi _v \sum\limits_{k = 1}^P {\kappa _v f_{pk}^2 D_{qi} \phi _{pk} C_{pk}^{exe} \frac{{\partial ^2 \varepsilon _{RSUi}^* }}{{\partial p_{pai}^2 }}}  \\ 
	\end{array} \right.
\end{equation}
\begin{equation}
	\left\{ \begin{array}{l}
		\frac{{\partial U_{RSUi} }}{{\partial p_{RSUi} }} = \varepsilon _{RSUi}^* D_{qi}  + D_{qi} p_{RSUi} \frac{{\partial \varepsilon _{RSUi}^* }}{{\partial p_{RSUi} }} \\ 
		\begin{array}{*{20}c}
			{} & {}  \\
		\end{array} - \xi _r \sum\limits_{j = 1}^R {\kappa _r f_{rk}^2 D_{qi} \phi _{rj} C_{rj}^{exe} \frac{{\partial \varepsilon _{RSUi}^* }}{{\partial p_{RSUi} }}}  \\ 
		\frac{{\partial ^2 U_{RSUi} }}{{\partial p_{RSUi}^2 }} = D_{qi} p_{RSUi} \frac{{\partial ^2 \varepsilon _{RSUi}^* }}{{\partial p_{RSUi}^2 }} + 2D_{qi} \frac{{\partial \varepsilon _{RSUi}^* }}{{\partial p_{RSUi} }} \\ 
		\begin{array}{*{20}c}
			{} & {}  \\
		\end{array} - \xi _r \sum\limits_{j = 1}^R {\kappa _r f_{rk}^2 D_{qi} \phi _{rj} C_{rj}^{exe} \frac{{\partial ^2 \varepsilon _{RSUi}^* }}{{\partial p_{RSUi}^2 }}}  \\ 
	\end{array} \right.
\end{equation}

Substituting equation (30) into the above two equations reveals that the second derivative of utility functions are less than zero when $T_{pai} \neq T_{RSUi}$. In the case of $T_{pai} = T_{RSUi}$, there exists a unique intersection between $U_{pai}$ and $U_{RSUi}$. As a result, the Nash equilibrium exists between RSU and PV.

In order to demonstrate the uniqueness of Nash equilibrium, the PV and RSU best response functions should be formulated. Given the pricing strategy of either PV or RSU, the best response functions for PV and RSU are defined as follows.
\begin{equation}
	\mathbb{F}(p_{pai} ) = \mathop {\arg \max }\limits_{p_{RSUi} } U_{RSUi} (p_{RSUi} ,p_{pai} )
\end{equation}
\begin{equation}
	\mathbb{F}(p_{RSUi} ) = \mathop {\arg \max }\limits_{p_{pai} } U_{pai} (p_{RSUi} ,p_{pai} )
\end{equation}

Let ($p_{RSUi}^*$, $p_{_{pai} }^* $) be the optimal strategy of the RSU and PV, then the following conditions should be satisfied.
\begin{equation}
	p_{RSUi}^*  = \mathbb{F}(p_{_{pai} }^* ), p_{_{pai} }^*  = \mathbb{F}(p_{_{RSUi} }^* )	
\end{equation}

Theorem 2: The Nash equilibrium of RSU and PV is unique.

When $T_{pai} =T_{RSUi}$, $\varepsilon _{RSUi}^*  = \frac{{\Gamma _{pai} }}{{\Gamma _{RSUi}  + \Gamma _{pai} }}$ is a constant, then the intersection point of $U_{pai}$ and $U_{RSUi}$ is the unique Nash equilibrium point, which can be calculated as 
\begin{equation}
	p_{RSUi}^*  = p_{pai}  \cdot \frac{{\Gamma _{RSUi} }}{{\Gamma _{pai} }} - \frac{{\Gamma _{RSUi}  + \Gamma _{pai} }}{{D_{qi} \Gamma _{pai} }} \cdot M_i 	
\end{equation}

\begin{equation}
	p_{_{pai} }^* {\rm{ = }}p_{RSUi}  \cdot \frac{{\Gamma _{RSUi} }}{{\Gamma _{pai} }} + \frac{{\Gamma _{RSUi}  + \Gamma _{pai} }}{{D_{qi} \Gamma _{RSUi} }} \cdot M_i 	
\end{equation}

Where 
\[\begin{array}{l}
	M_i  = \xi _v \left( {\frac{{\Gamma _{RSUi} }}{{\Gamma _{RSUi}  + \Gamma _{pai} }} \cdot \sum\limits_{k = 1}^P {\kappa _v D_{qi} \phi _{pk} f_{pk}^2  + } \frac{1}{{D_{Bv} /\varpi }}E_{pa}^{BC} } \right) \\ 
	- \xi _v \left( {\frac{{\Gamma _{RSUi} }}{{\Gamma _{RSUi}  + \Gamma _{pai} }} \cdot \sum\limits_{j = 1}^R {\kappa _r D_{qi} \phi _{rj} f_{rj}^2  + } \frac{1}{{D_{Bv} /\varpi }}E_{RSU}^{BC} } \right) \\ 
\end{array}
\]

When $T_{pai} \neq T_{RSUi}$, it can be proved that $\mathbb{F}(p_{pai} )$ and $\mathbb{F}(p_{RSUi} )$ are standard functions, therefore, the Nash equilibrium of RSU and PV is unique.

\subsection{BPVEC Offloading Strategy Algorithm}
To optimize the offloading strategy for RVs and the pricing strategies for RSUs and PVs, we design a BPVEC offloading strategy algorithm based on the gradient descent method, as shown in Algorithm 2.
\begin{algorithm}[H]
	\caption{BPVEC offloading strategy algorithm based on gradient descent method}
	\begin{algorithmic}[1]
		\STATE \textbf{Input:} set of RVs $\mathcal{V}$=$\{$1,2,...,$V$$\}$;		
		\STATE \textbf{Output:} offloading strategy $\varepsilon _{RSUi}^*$ , optimal pricing $p_{_{pai} }^*$, $p_{RSUi}^*$;
		\STATE \textbf{Initialization:} $\varepsilon _{RSUi}=0.5,p_{pai} =0.2,p_{RSUi}=0.5$;
		\STATE $ T_{pai},T_{RSUi}\gets$ equation (9) and (10)	
		\FOR{$ i=1:V$ }
		\REPEAT
		\STATE According to  $ T_{pai}$ and $T_{RSUi}$, $\varepsilon _{RSUi}\gets$equation (30)
		\REPEAT
		\IF{$T_{pai} \neq T_{RSUi}$}
		\STATE $ \nabla U_{pai}(t) \gets$ equation (31)
		\IF{$U_{pai}(t) \leq 0$}
		\STATE $p_{pai} (t+1)=p_{pai} (t)/\omega_{1}$
		\ELSE
		\STATE $p_{pai} (t+1)=p_{pai} (t)+\mu_{1} \nabla U_{pai }(t) $
		\ENDIF
		\ELSE
		\IF{$U_{pai} (t) <0 $} 
		\STATE $p_{pai} (t+1)=p_{pai }(t)/\omega_1$
		\ELSE
		\STATE $ p_{pai }(t+1) \gets $equation (36) 
		\ENDIF
		\ENDIF
		\STATE  update $U_{pai }(t) $
		\UNTIL{$\frac{{\left( {p_{pai} (t + 1) - p_{pai} (t)} \right)^2 }}{{(p_{pai} (t))^2 }} < \vartheta $}
		\REPEAT
		\IF {$T_{pai} \neq T_{RSUi}$}
		\STATE $ \nabla U_{RSUi}(t) \gets$ equation (32)
		\IF{$U_{RSUi}(t) \leq 0$}
		\STATE $p_{RSUi} (t+1)=p_{RSUi} (t)/\omega_{2}$
		\ELSE
		\STATE $p_{RSUi} (t+1)=p_{RSUi} (t)+\mu_{2} \nabla U_{RSUi }(t) $
		\ENDIF
		\ELSE
		\IF{$U_{RSUi} (t) <0 $} 
		\STATE $p_{RSUi} (t+1)=p_{RSUi }(t)/\omega_2$
		\ELSE
		\STATE $ p_{RSUi }(t+1) \gets $equation (37) 
		\ENDIF
		\ENDIF
		\STATE  update $U_{RSUi }(t) $
		\UNTIL{$\frac{{\left( {p_{RSUi} (t + 1) - p_{RSUi} (t)} \right)^2 }}{{(p_{RSUi} (t))^2 }} < \vartheta$}
		\UNTIL{$\frac{{\left( {\varepsilon _{RSUi} (t + 1) - \varepsilon _{RSUi} (t)} \right)^2 }}{{(\varepsilon _{RSUi} (t))^2 }} < \vartheta$}
		\ENDFOR
		\label{code:BPVEC}
	\end{algorithmic}
\end{algorithm}

In Algorithm 2, $\nabla U_{pai }$ and $\nabla U_{RSUi}$ are respectively the gradient with $\frac{{\partial U_{pai} }}{{\partial p_{pai} }}$ and $\frac{{\partial U_{RSUi} }}{{\partial p_{RSUi} }}$, which calculated by equation (31) and (32). $\vartheta$ is the iteration termination condition variation with minimum value. $\mu_1$ and $\mu_2$ are denoted for the gradient descent learning rate. The Algorithm 2 aims to obtain the offloading strategy $\varepsilon _{RSUi}$ and the optimal pricing $p_{pai},p_{RSUi}$. After calculating the $T_{pai}$ and $T_{RSUi}$ by equation (9) and (10), two situations should be considered, i.e., $T_{pai} \neq T_{RSUi}$ and $T_{pai} =T_{RSUi}$ ,  as depicted in lines 7,12 and lines 21,26. When $T_{pai} \neq T_{RSUi}$, the optimal values are obtain from the gradient descent method, as shown in lines 10 and 24. Otherwise, they should be calculated by equation (36) and (37). Assuming the algorithm requires $M$ iterations to satisfy the loop termination condition, Algorithm 2 involves computing one gradient per iteration, thus resulting in a computational complexity of $O(M)$. When considering the processing of $V$ RVs, the overall computational complexity scales to $O(VM)$.
\section{Experimental Results and Analysis}
To evaluate the performance of proposed scheme, Matlab is used for simulation, and the primary simulation parameters are detailed in Table~\ref{tab:parameters}.
\begin{table}[H]
	\caption{The main simulation parameters}
	\label{tab:parameters}
	\centering
	\begin{tabular}{|l|l|l|l|}
		\hline
		Parameters & Value & Parameters & Value\\
		\hline
		$C^{exe}_{pk}, C^{exe}_{rj}$ & 24cycles/bit \cite{LangTian-597} & $T_{maxi}$ & [100,200]ms\\
		\hline
		$d_0$ & 100m \cite{MaZhao-611} & $W_b$ & 15MB\\
		\hline
		$D_{qi}$ & [10, 30]MB &	$\alpha$ & 1 \cite{SuXu-613}\\
		\hline
		$D_{Bv}$ & 4MB & $\beta$ & $10^6$ cycles \cite{LangTian-597} \\
		\hline
		$f_{pk}$ & [1, 2.5]GHz & $\delta$ & 2 \cite{MaZhao-611}\\
		\hline
		$f_{rj}$ & [4, 6]GHz & $\eta$ & 1.63726×$10^{-9}$ \cite{MaZhao-611}\\
		\hline
		$N_0$ &	1.2589×$10^{-13}$W \cite{MaZhao-611} & $\theta$ & 10×$10^6$cycles \cite{LangTian-597}\\
		\hline
		$P_t$ &	0.28183815W \cite{MaZhao-611} & $\kappa_v$ & $10^{-27}$ \cite{FengYu-614}\\
		\hline
		$p_{pai} , p_{RSUi}$ & [0.1,$\infty $] & $\kappa_r$ &	$10^{-28}$\\
		\hline
		$p_{th}$ & 	0.95 & $\varpi$ & 1KB\\
		\hline
	\end{tabular}
\end{table}

\begin{figure*}[htbp]
	\centering
	\subfigure[]{
		\begin{minipage}[t]{0.48\textwidth}
			\centering
			\includegraphics[scale=0.3]{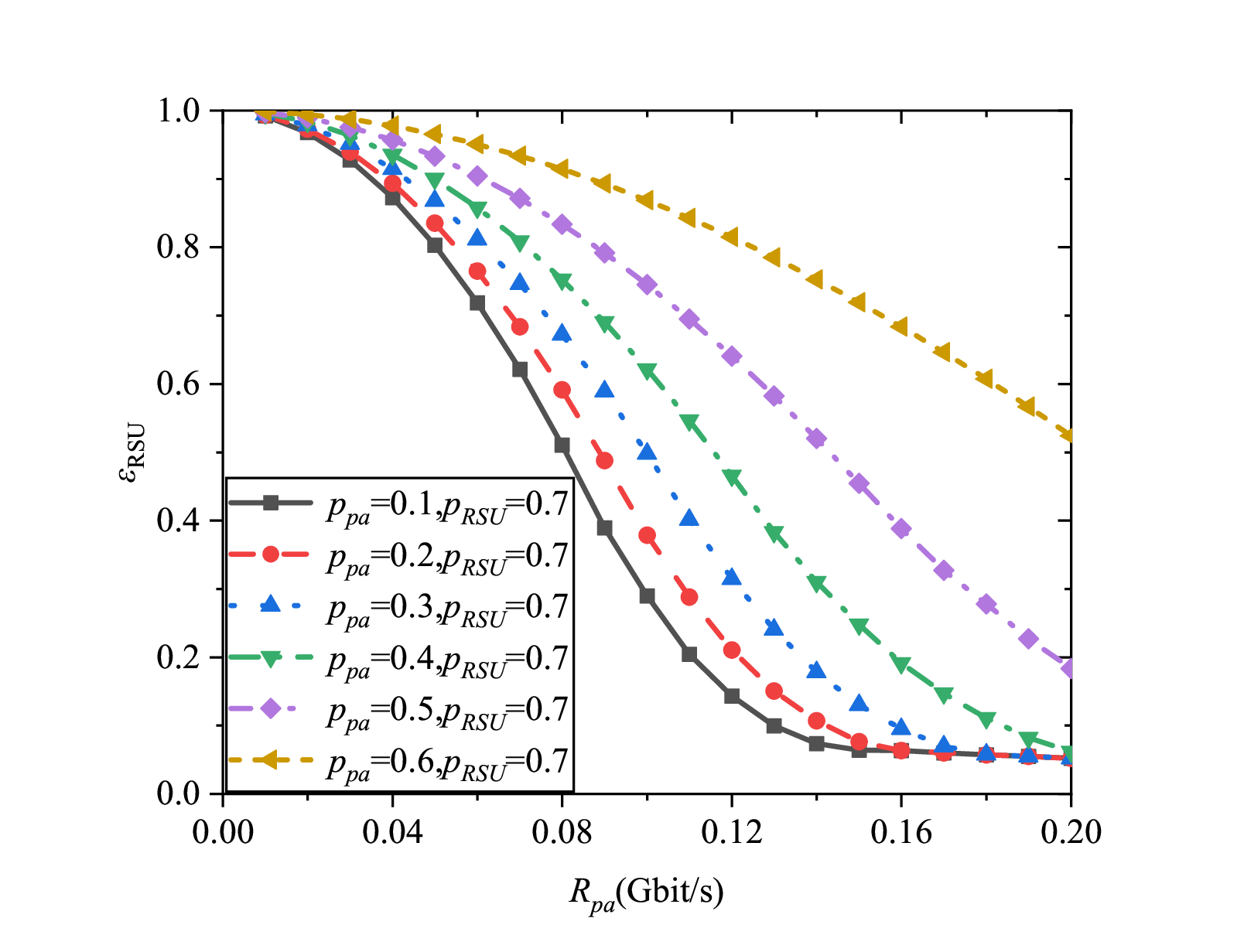}
		\end{minipage}
	}
	\hfill
	\subfigure[]{
		\begin{minipage}[t]{0.48\textwidth}
			\centering
			\includegraphics[scale=0.3]{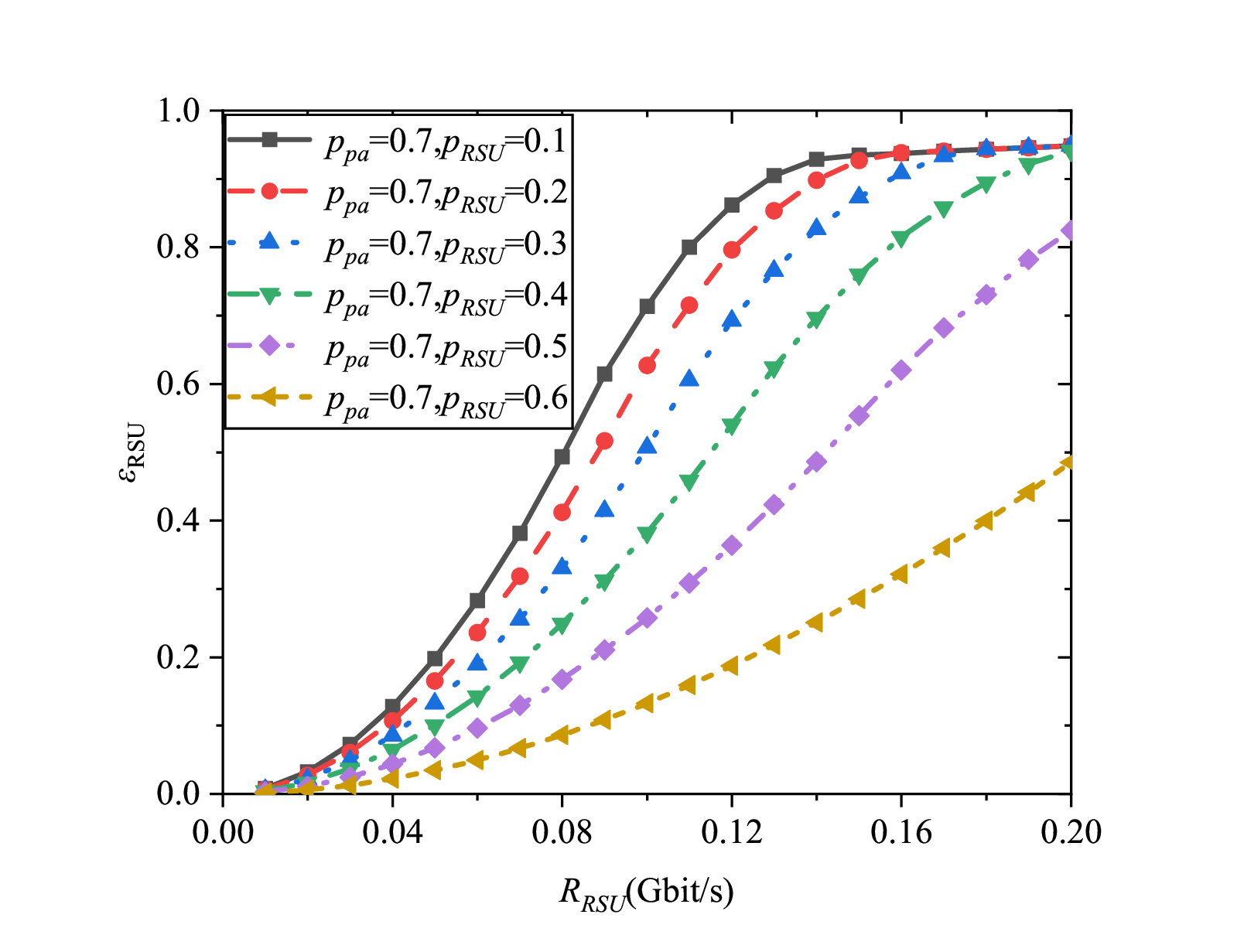}
		\end{minipage}
	}
	\centering
	\caption{Variation of the average  $\varepsilon _{RSUi}$ with the transmission rate. (a) PV transmission rate $R_{pa}$. (b) RSU transmission rate $R_{RSU}$.}
	\label{fig3:transmission}
\end{figure*}

Firstly, the variation of the average $\varepsilon _{RSUi}$ of the offloading strategy with the transmission rate is examined, as depicted in Fig.~\ref{fig3:transmission}. In Fig.~\ref{fig3:transmission}(a), the $\varepsilon _{RSUi}$ decreases with the increase of the transmission rate $R_{pa}$ from the RV to the PV, but increases with higher pricing of PV. Fig.~\ref{fig3:transmission}(b) illustrates that $\varepsilon _{RSUi}$ rises as the transmission rate $R_{RSU}$ of the RSU increases, and gradually decreases with the rise in pricing $p_{RSU}$. It indicates that users tend to offload computing tasks to the entity offering more favorable pricing, and a higher transmission rate enhances the willingness of RVs to offload.

\begin{figure*}[htbp]
	\centering
	\subfigure[]{
		\begin{minipage}[t]{0.48\textwidth}
			\centering
			\includegraphics[scale=0.3]{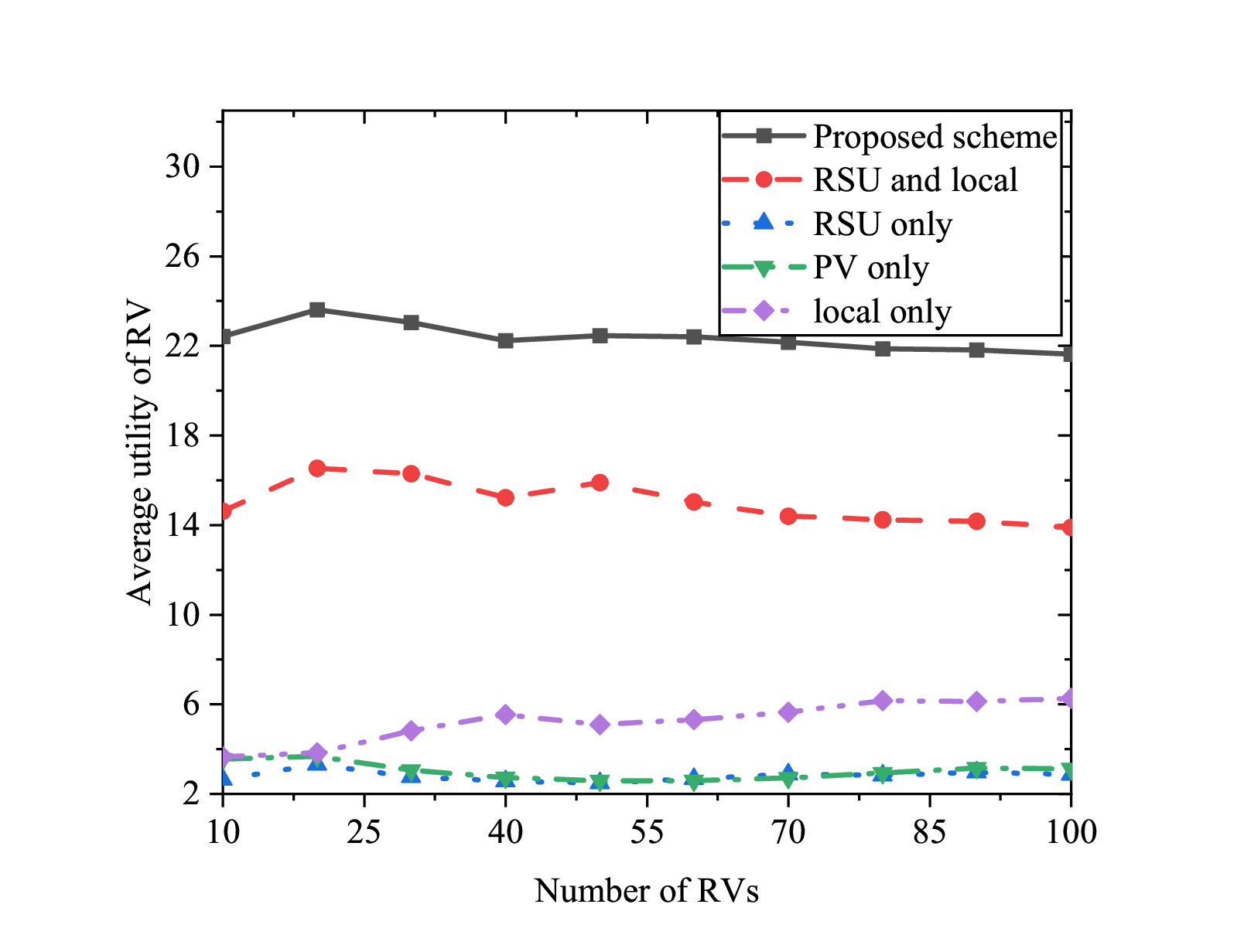}
		\end{minipage}
	}
	\hfill
	\subfigure[]{
		\begin{minipage}[t]{0.48\textwidth}
			\centering
			\includegraphics[scale=0.3]{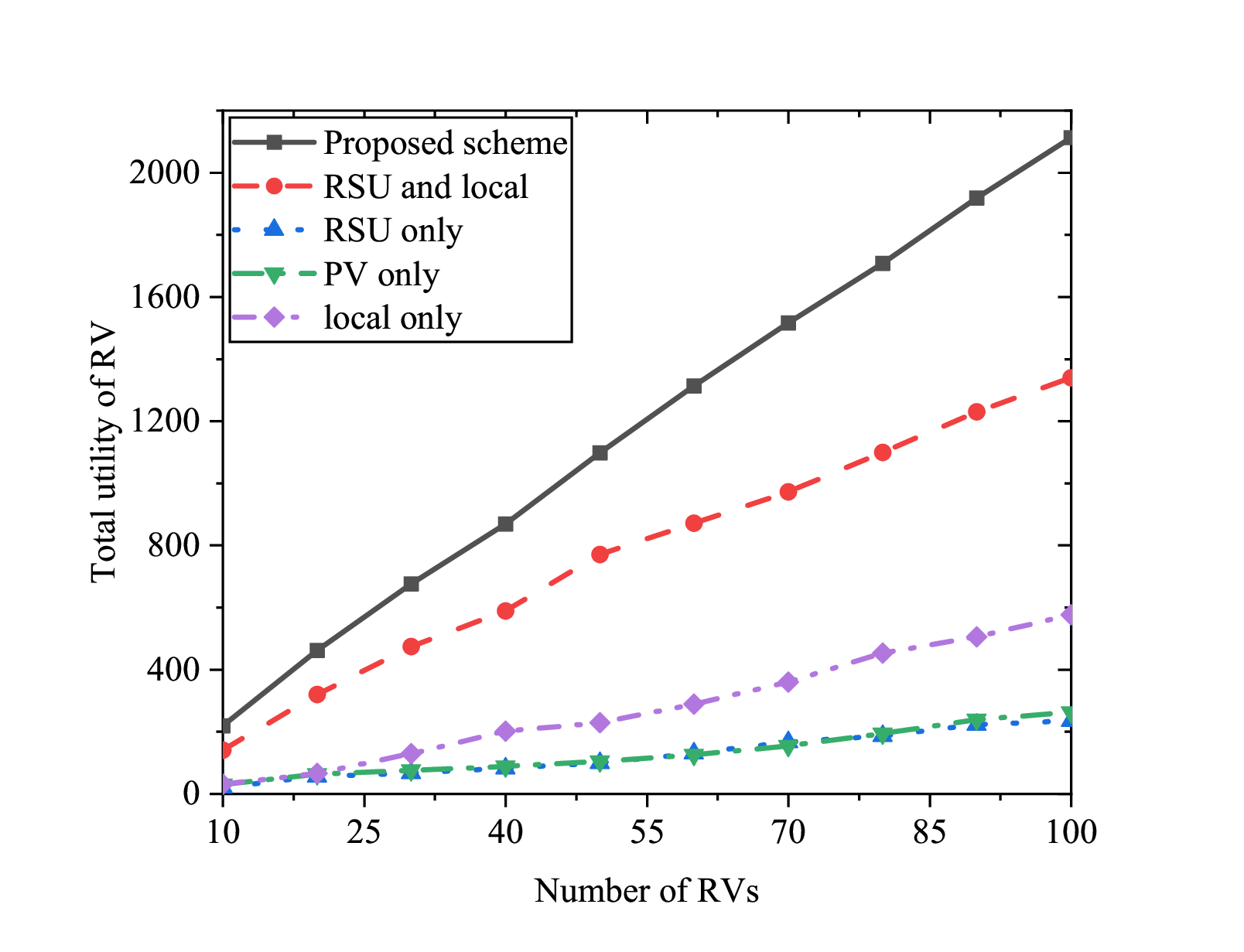}
		\end{minipage}
	}
	\centering
	\caption{Variation of RV utility function with RV number in different offloading schemes. (a) Average RV utility. (b) Total RV utility.}
	\label{fig4:offloading schemes}
\end{figure*}

Secondly, the performance of proposed scheme is compared with traditional offloading schemes, as illustrated in Fig.~\ref{fig4:offloading schemes}. In the RSU and local scenario, the RV can execute local computation or offload tasks to the RSU for processing. The RSU only and PV only scenarios involve the RV offloading all computation tasks to the RSU or the PV, respectively. The local only signifies that the RV performs only local computation. The figure demonstrates that sharing computation resources between the RSU and PV significantly alleviates the RV’s computational load, enabling tasks to be completed more rapidly and thus reducing the RV’s computing costs and enhancing its utility. Therefore, the proposed scheme outperforms other schemes for RV utility.

%
%

\begin{figure*}[htbp]
	\centering
	\subfigure[]{
		\begin{minipage}[t]{0.3\textwidth}
			\includegraphics[width=\textwidth]{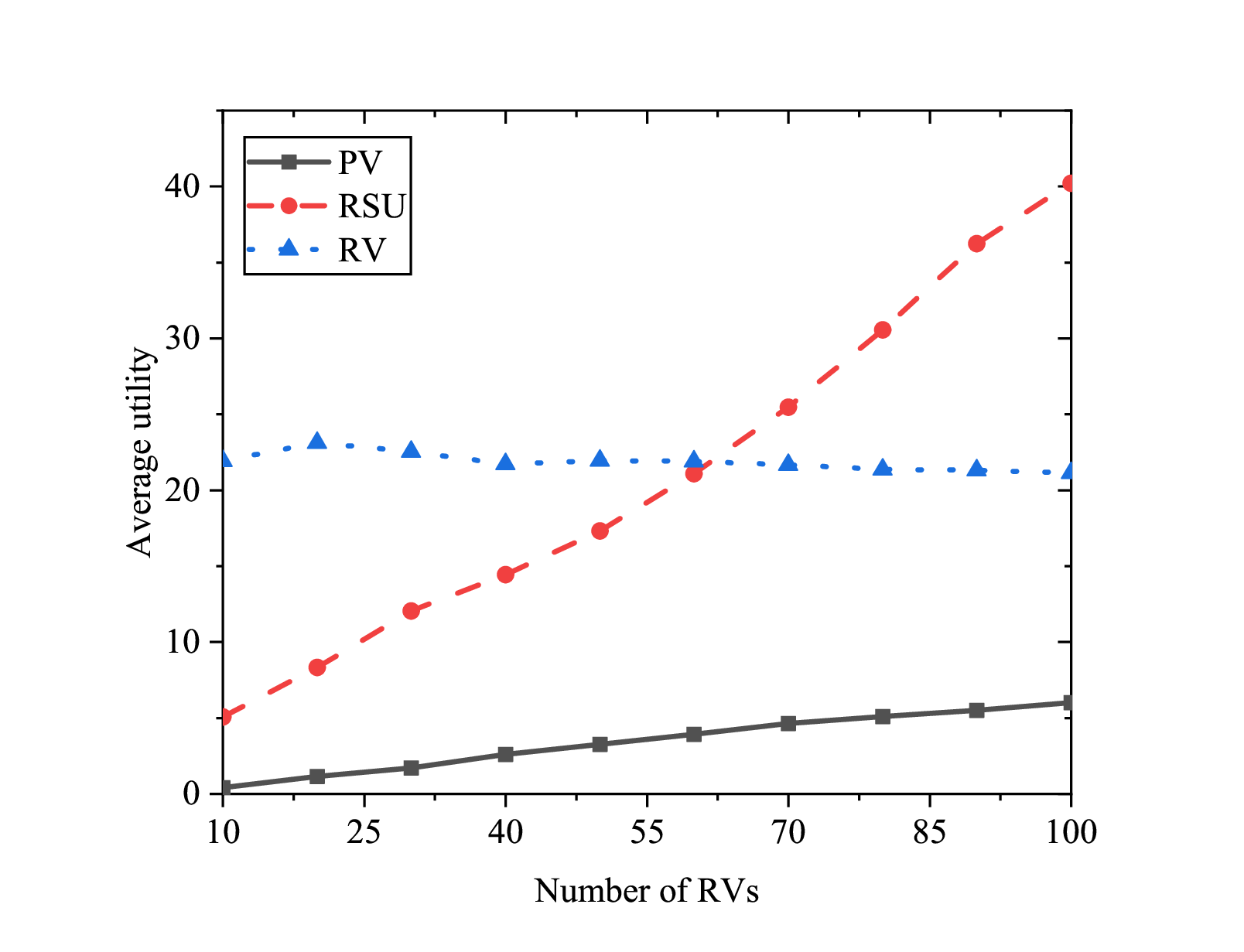}
		\end{minipage}
	}
	\hfill
	\subfigure[]{
		\begin{minipage}[t]{0.3\textwidth}
			\includegraphics[width=\textwidth]{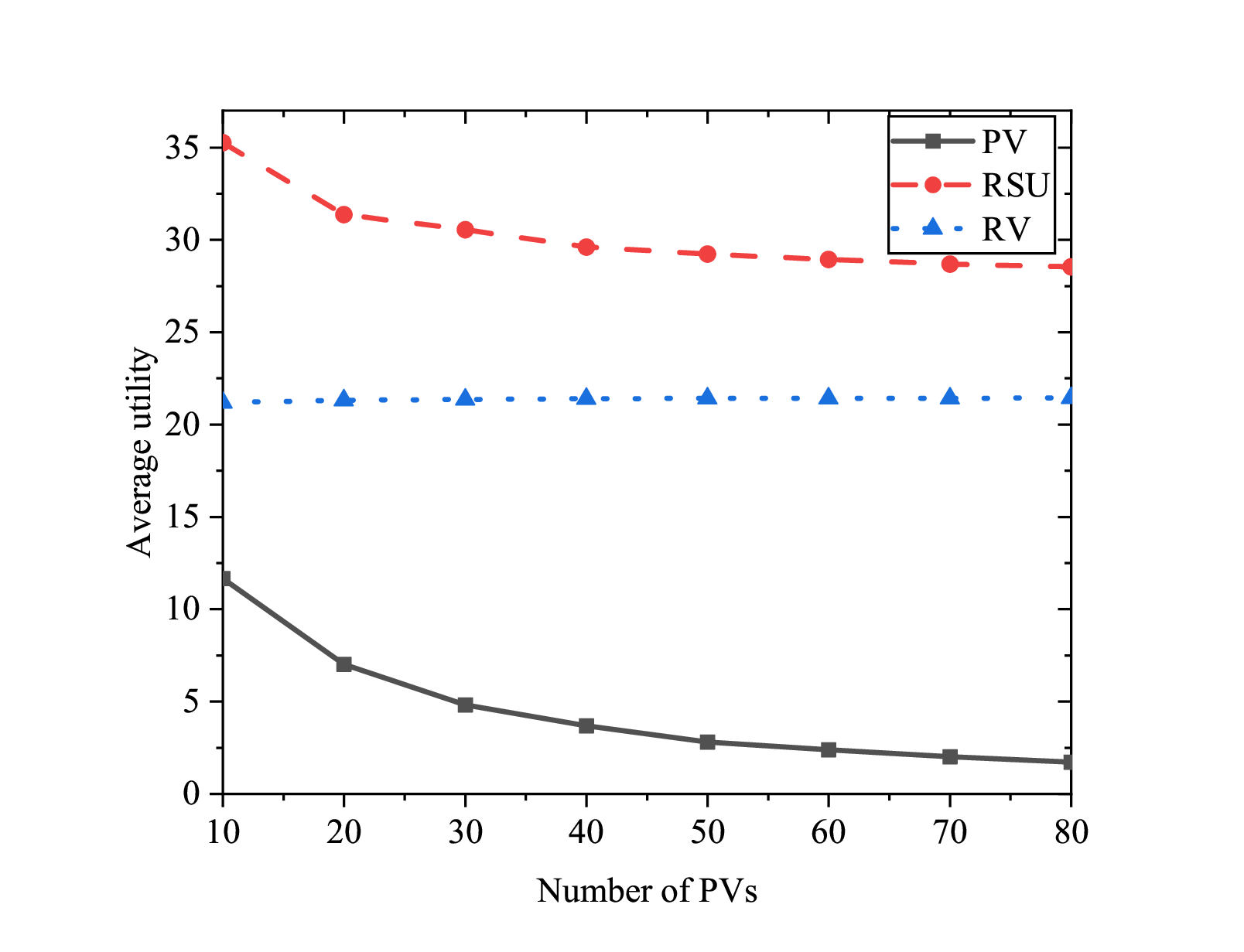}
		\end{minipage}
	}
	\hfill
	\subfigure[]{
		\begin{minipage}[t]{0.3\textwidth}
			\includegraphics[width=\textwidth]{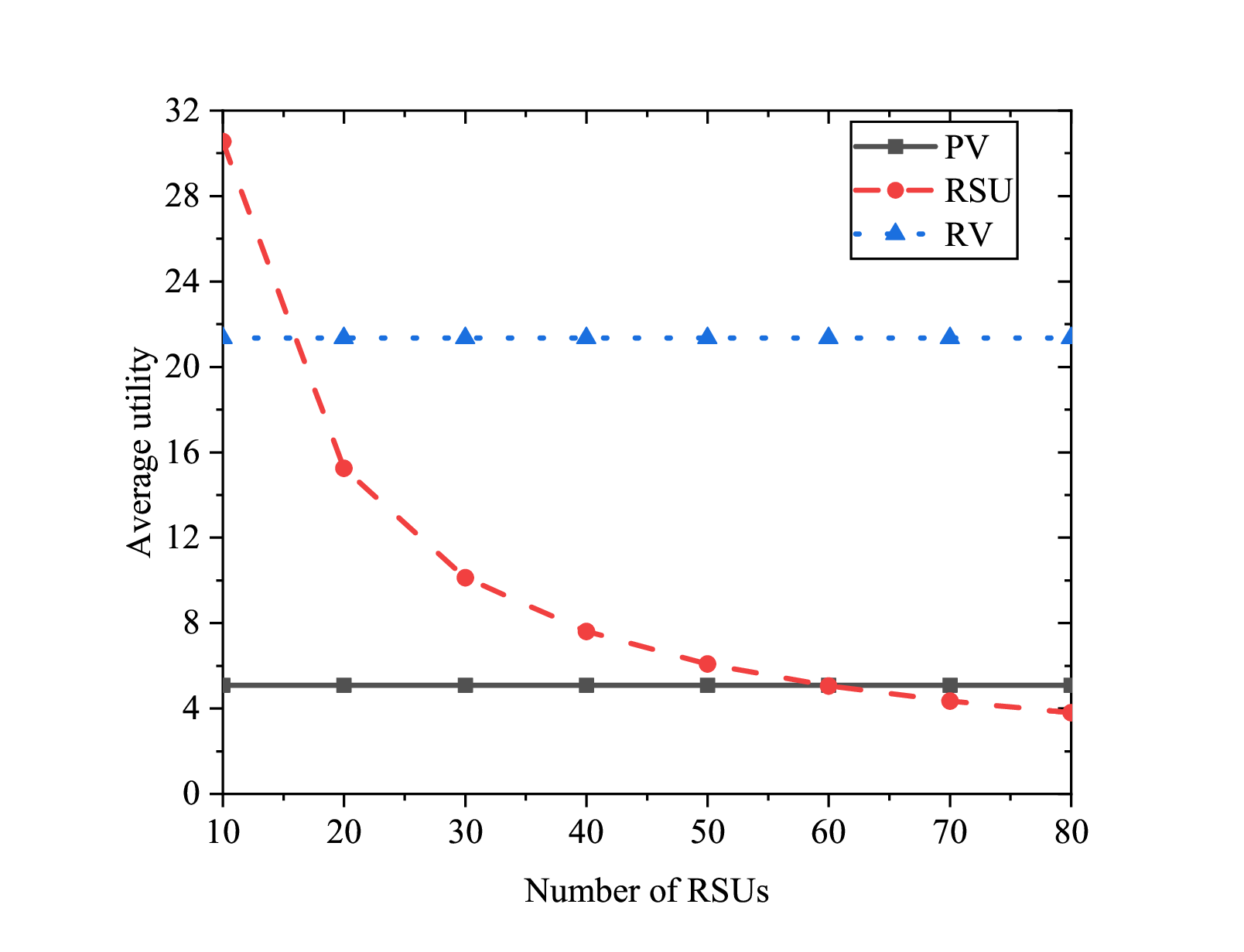}
		\end{minipage}
	}
	\caption{Effect of different RVs, PVs and RSUs on utility function. (a) RVs. (b) PVs. (c) RSUs.}
	\label{fig5:different RVs}
\end{figure*}

Next, the impact of varying numbers of PVs, RSUs, and RVs on the average utility function values is analyzed, as depicted in Fig.~\ref{fig5:different RVs}. Fig.~\ref{fig5:different RVs}(a) shows the change of the average utility function with respect to the number of RVs. As the number of RVs increases, the overall computational task rises, which increases the task of offloading to PVs and RSUs, resulting in an increase in their utility. However, the utility change per individual RV is minimal. Fig.~\ref{fig5:different RVs}(b) and Fig.~\ref{fig5:different RVs}(c) illustrate the average utility function changes with respect to the number of RSUs and PVs, respectively. As the number of PVs and RSUs increases, the average utility function of PVs and RSUs decreases respectively. The reason is that the total tasks that the RV offloads to PVs and RSUs remain constant, the average benefit of each PV or RSU decreases.

%

\begin{figure*}[htbp]
	\centering
	\subfigure[]{
		\begin{minipage}[t]{0.48\textwidth}
			\centering
			\includegraphics[scale=0.3]{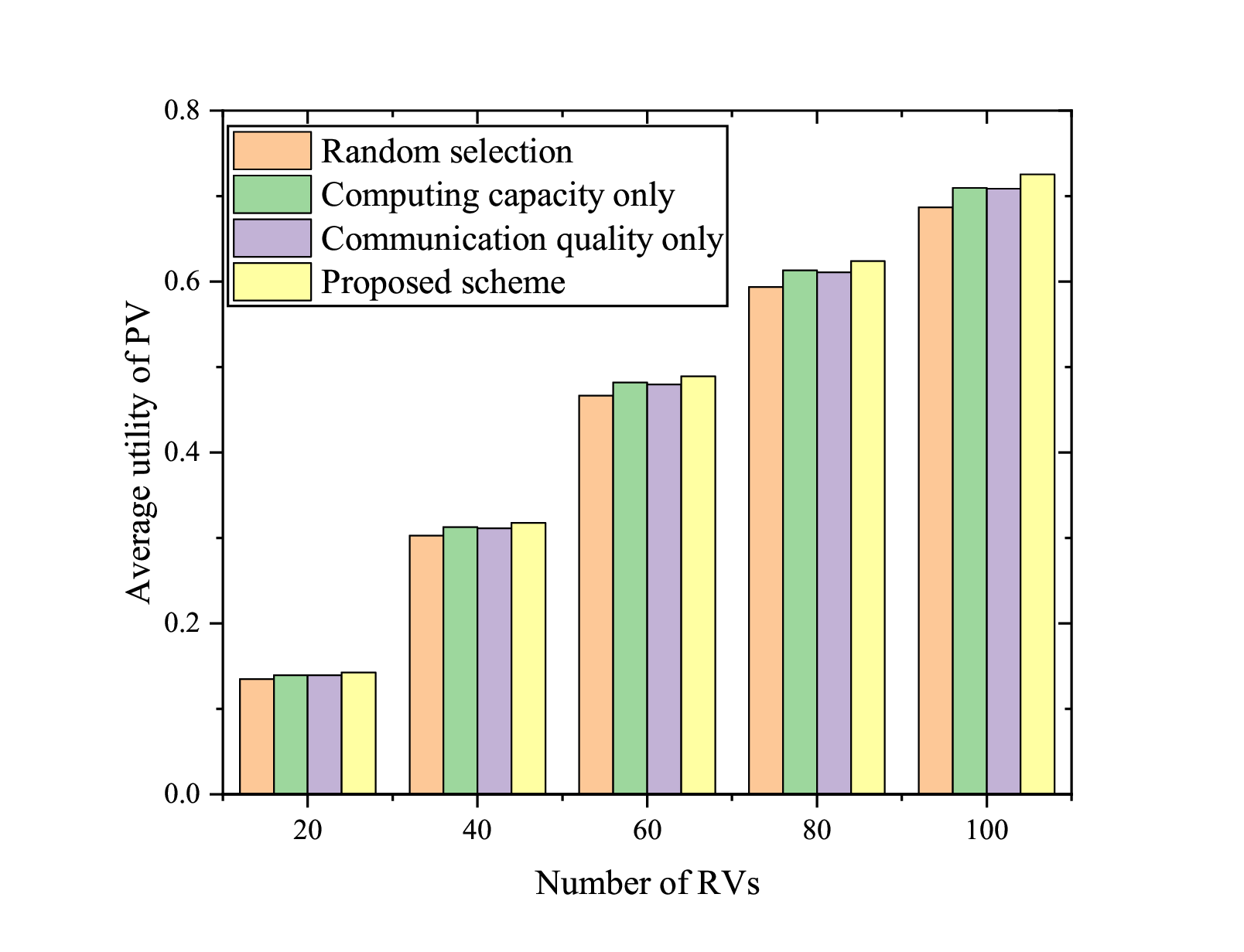}
		\end{minipage}
	}
	\hfill
	\subfigure[]{
		\begin{minipage}[t]{0.48\textwidth}
			\centering
			\includegraphics[scale=0.3]{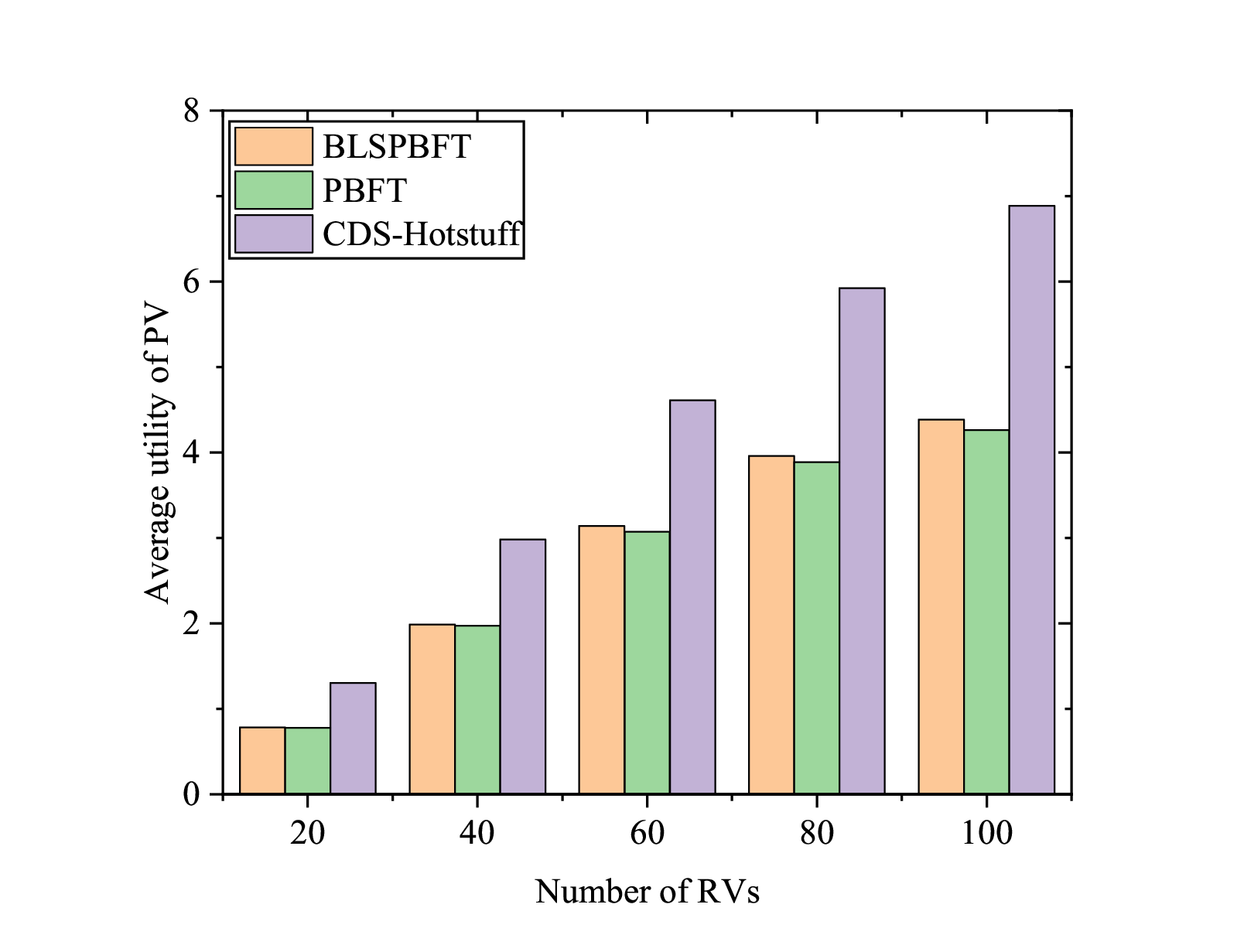}
		\end{minipage}
	}
	\centering
	\caption{Compare with different node selection strategies and consensus schemes. (a) Different node selection strategies. (b) Different consensus schemes.}
	\label{fig6:selection strategy and consensus schemes}	
\end{figure*}

Then, to assess the performance of the proposed consensus scheme, different node selection strategies and consensus schemes are compared in Fig.~\ref{fig6:selection strategy and consensus schemes}. Fig.~\ref{fig6:selection strategy and consensus schemes}(a) compares different node selection strategies, where random selection, computing capacity only, and communication quality only represent HotStuff's methods for selecting consensus nodes based on random choice, computational capacity, and communication quality, respectively. The figure shows that the proposed scheme outperforms other schemes in terms of the utility function. This advantage is attributed to the strategic selection of nodes with superior computational capacity and signal-to-noise ratio, effectively reducing the energy consumption required for consensus. In addition, the selected consensus nodes have longer parking duration and better communication quality, which significantly enhances the reliability of the nodes. Fig.~\ref{fig6:selection strategy and consensus schemes}(b) compares various consensus schemes with the proposed scheme, where CDS-Hotstuff is the consensus scheme proposed in this paper, PBFT is the traditional Byzantine consensus mechanism, and BLSPBFT is an improved PBFT using BLS aggregate signature. CDS-Hotstuff generates less additional energy consumption in the consensus process compared to PBFT and BLSPBFT, which reduces the system energy consumption. Consequently, the average utility of CDS-Hotstuff is greater than BLSPBFT and PBFT.

\begin{figure*}[htbp]
	\centering
	\subfigure[]{
		\begin{minipage}[t]{0.48\textwidth}
			\centering
			\includegraphics[scale=0.3]{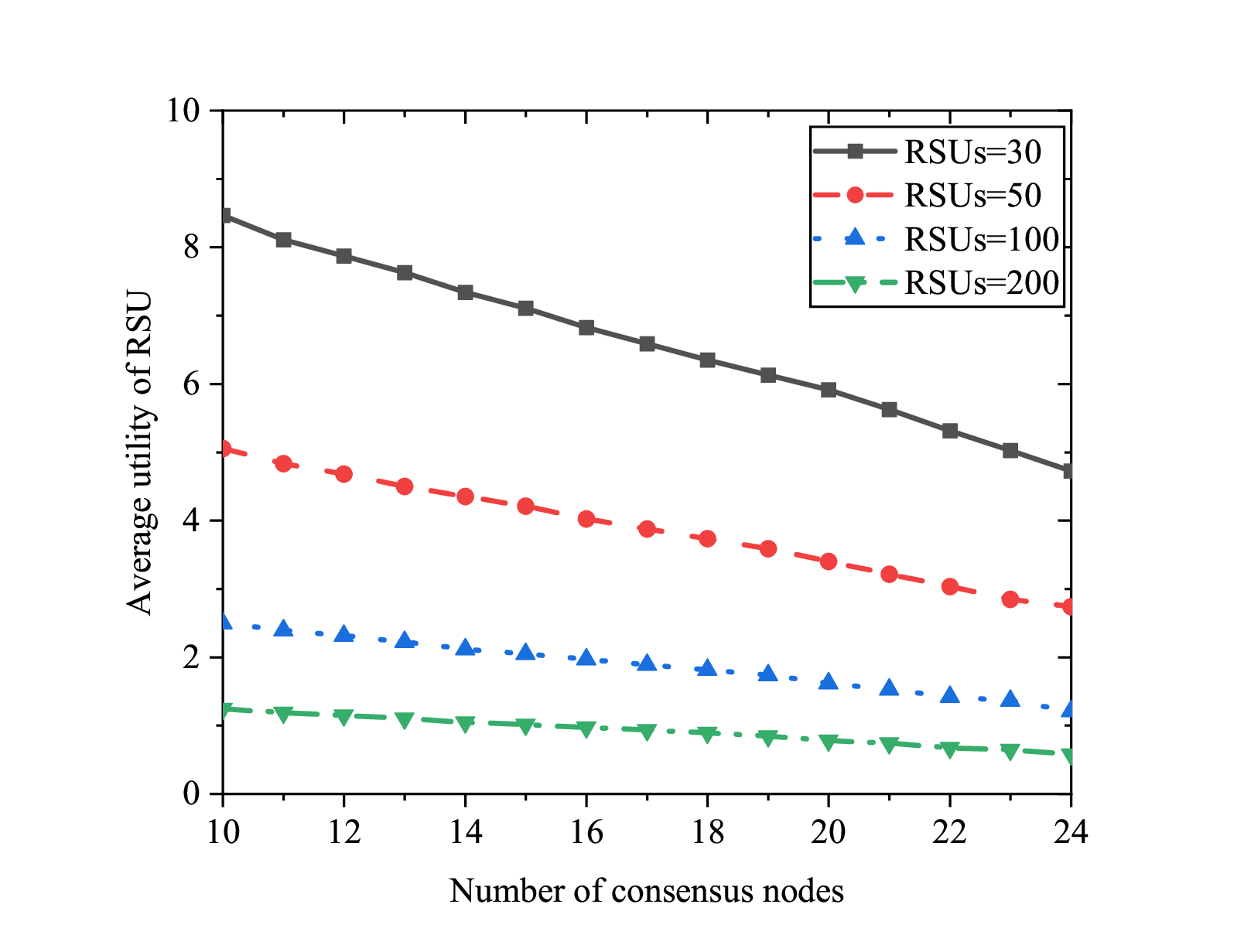}
		\end{minipage}
	}
	\hfill
	\subfigure[]{
		\begin{minipage}[t]{0.48\textwidth}
			\centering
			\includegraphics[scale=0.3]{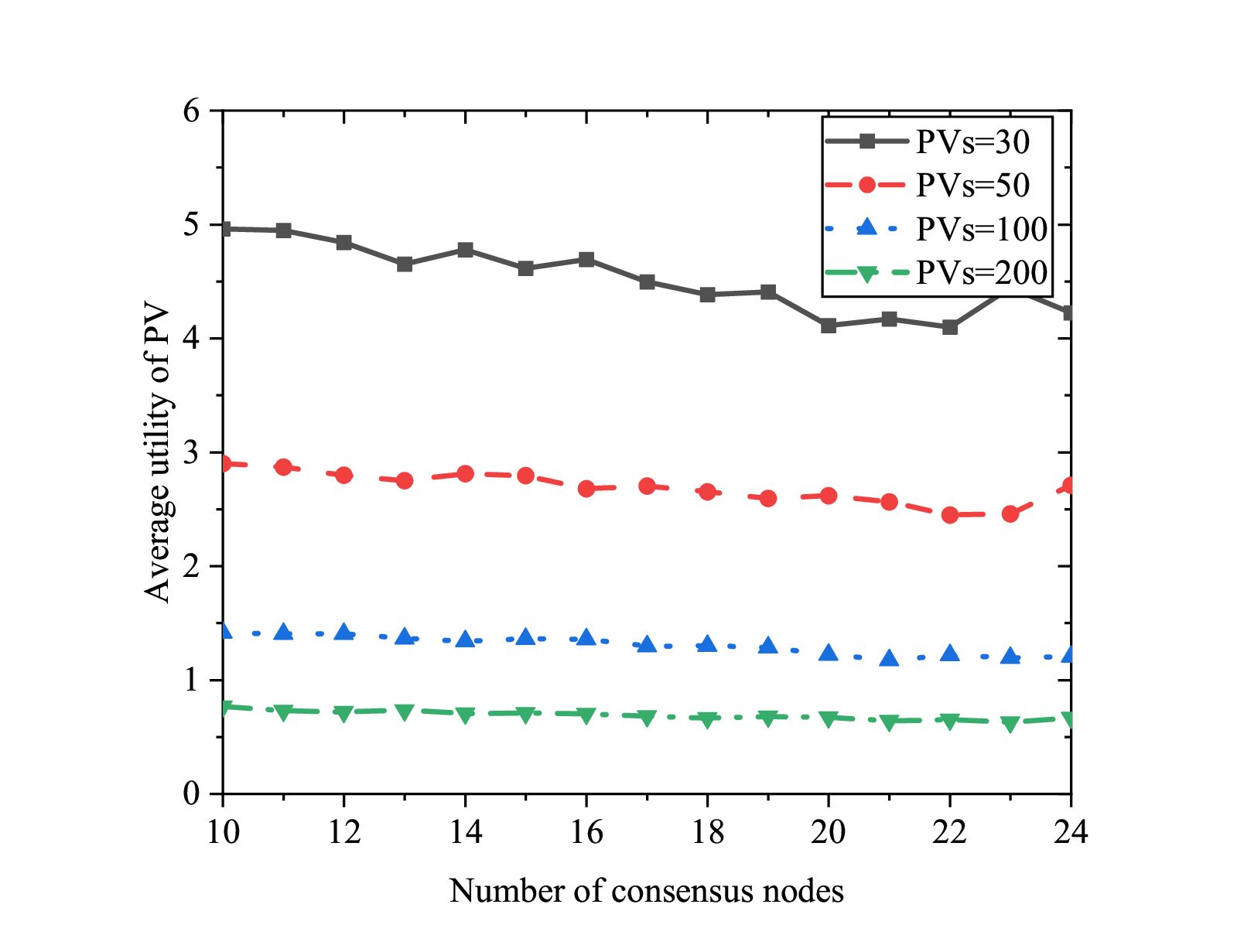}
		\end{minipage}
	}
	\centering
	\caption{Variation of utility function with consensus nodes. (a) RSU utility function. (b) PV utility function.}
	\label{fig7:consensus nodes}
\end{figure*}

Finally, we study the impact of varying the number of consensus nodes in CDS-Hotstuff on the utility function. As shown in Fig.~\ref{fig7:consensus nodes}, an increase in consensus nodes leads to a decline in the utility function for both RSUs and PVs. This occurs because adding more consensus nodes increases the energy consumption of the consensus mechanism, reducing the revenue and utility of RSUs and PVs.

\section{Conclusion}
In this paper, we introduce a BPVEC offloading framework that ensures the security and reliability of PVEC task offloading and transactions. To achieve this, we improve the Hotstuff consensus mechanism, optimizing the selection of parking nodes as consensus nodes according to parking time, computing capability and communication quality. This improvement aims to enhance the reliability of blockchain during the computation offloading process. Building on this, a Stackelberg game model with the RSU and PV as leaders and the RV as the follower is established, and the game model is analyzed by two-stage analysis and backward inductive method. Furthermore, an offloading strategy algorithm of BPVEC using gradient descent method is designed to maximize the system benefits. The simulation results indicate that our scheme has the advantages of safety and reliability while ensuring the maximum benefit. In future work, we will further study the computing offloading privacy issue, explore how to protect vehicle privacy during RV offloading in a distributed environment, and further ensure the security of RV offloading.

\bibliographystyle{IEEEtran}
\bibliography{IEEEabrv, article-cite-TVT}

\begin{IEEEbiography}[{\includegraphics[width=1in,height=1.25in,clip,keepaspectratio]{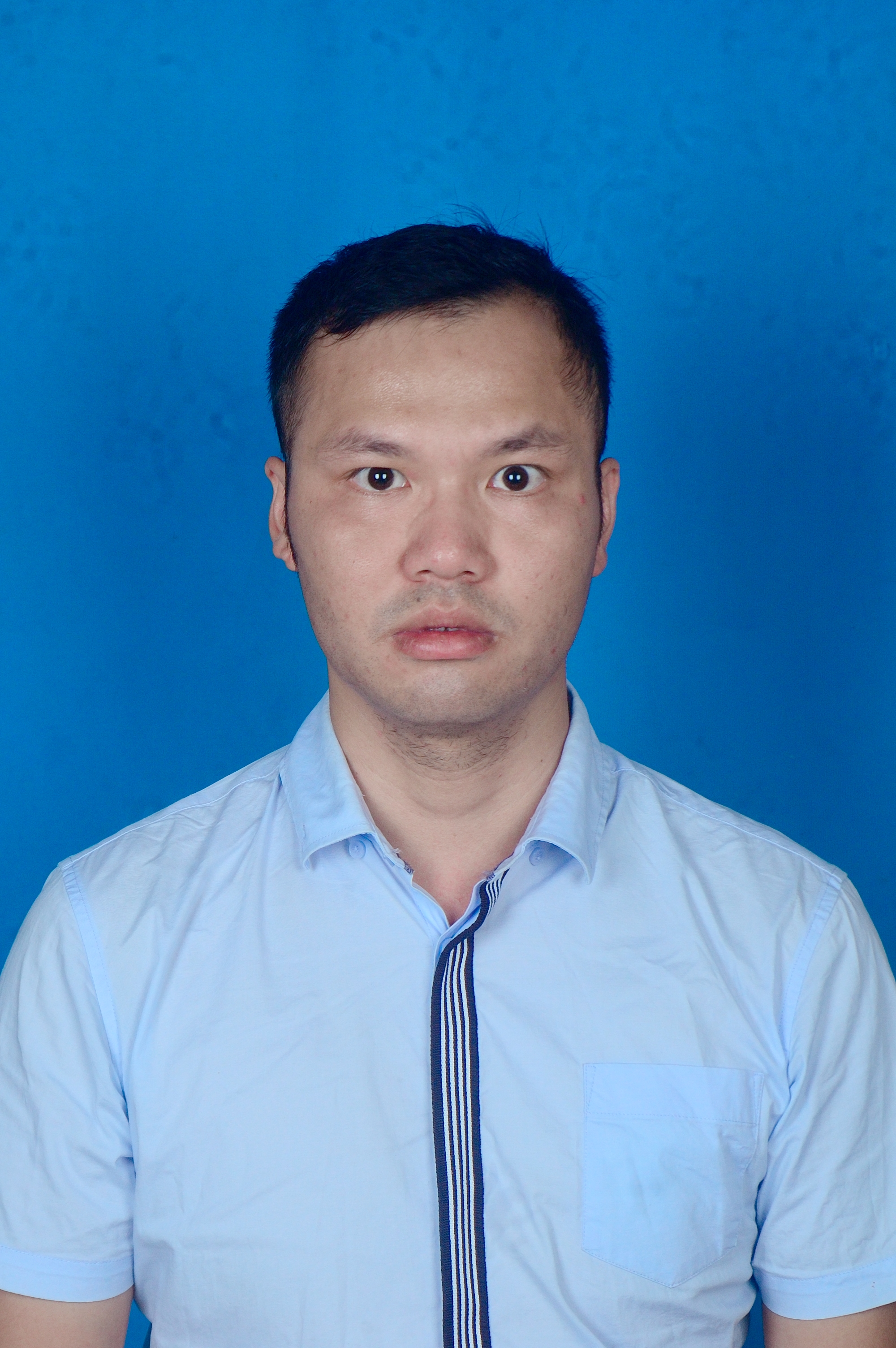}}]{Guoling Liang}
	received his Bachelor's and Master's degrees from Guilin University of Electronic Technology in China in 2012 and 2015, respectively, where he is currently pursuing the doctoral degree. He is currently a teacher in the School of Physics and Telecommunication Engineering at Yulin Normal University. His current research interests include wireless blockchain and edge computing.
\end{IEEEbiography}

\begin{IEEEbiography}[{\includegraphics[width=1in,height=1.25in,clip,keepaspectratio]{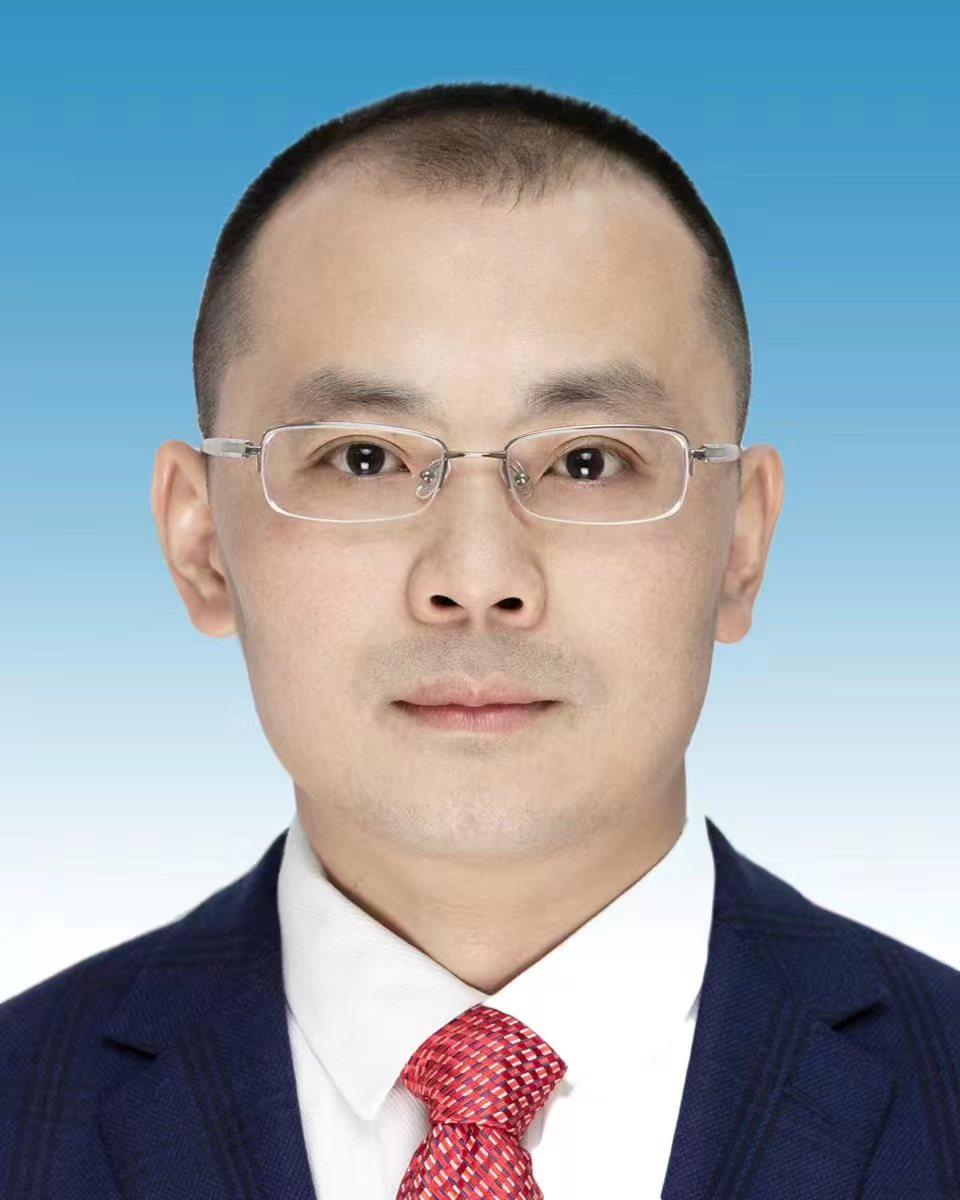}}]{Chunhai Li}
	received the Ph.D. degree from Guilin University of Electronic Technology, Guilin, China, in 2020. He is a Professor with the Guangxi Engineering Research Center of Industrial Internet Security and Blockchain, Guilin University of Electronic Technology. His current research interests include Internet of things security, network security and blockchain.
\end{IEEEbiography}

\begin{IEEEbiography}[{\includegraphics[width=1in,height=1.25in,clip,keepaspectratio]{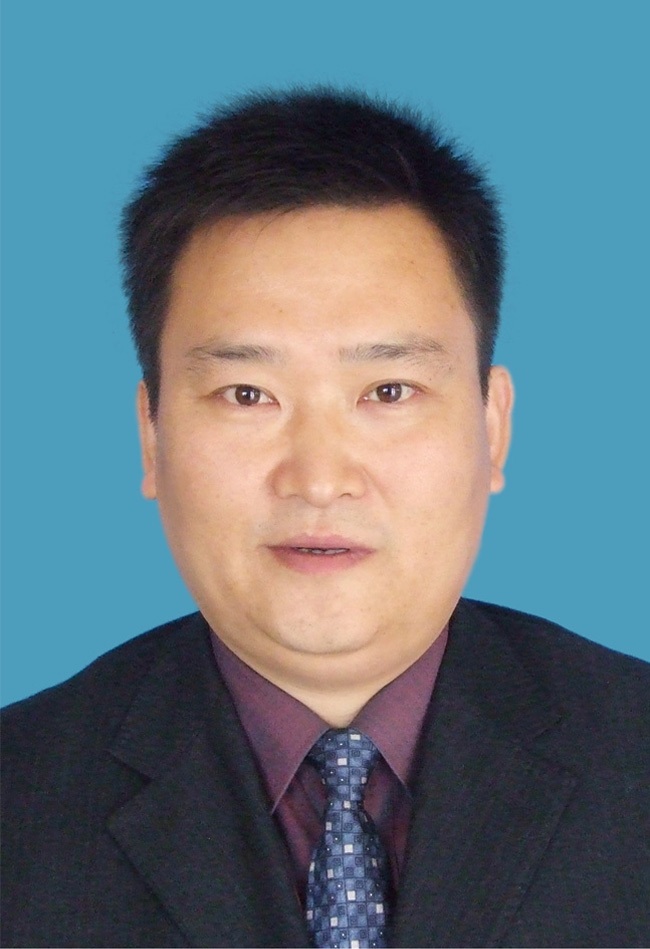}}]{Feng Zhao}
	received the Ph.D. degree in communication and information systems from Shandong University, China, in 2007. He is currently a Professor with the Guangxi Engineering Research Center of Industrial Internet Security and Blockchain, Guilin University of Electronic Technology, Guilin, China. His research interests include cognitive radio networks, MIMO technologies, cooperative communications, and information security.
\end{IEEEbiography}

\begin{IEEEbiography}[{\includegraphics[width=1in,height=1.25in,clip,keepaspectratio]{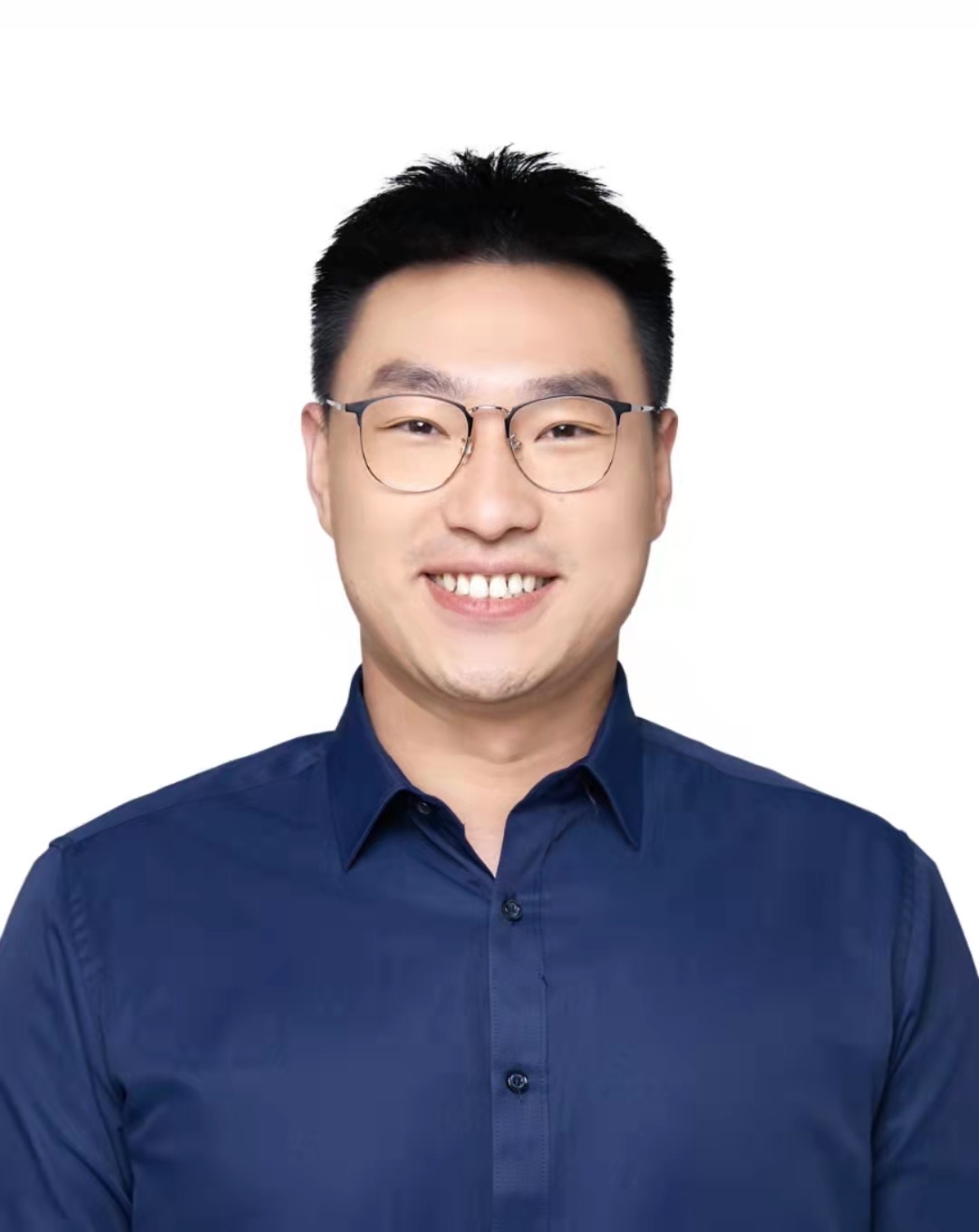}}]{Chuan Zhang}
	received his Ph.D. degree in computer science from Beijing Institute of Technology, Beijing, China, in 2021. From Sept. 2019 to Sept. 2020, he worked as a visiting Ph.D. student with the BBCR Group, Department of Electrical and Computer Engineering, University of Waterloo, Canada. He is currently an assistant professor at the School of Cyberspace Science and Technology, Beijing Institute of Technology. His research interests include secure data services in cloud computing, applied cryptography, machine learning, and blockchain.
\end{IEEEbiography}

\begin{IEEEbiography}[{\includegraphics[width=1in,height=1.25in,clip,keepaspectratio]{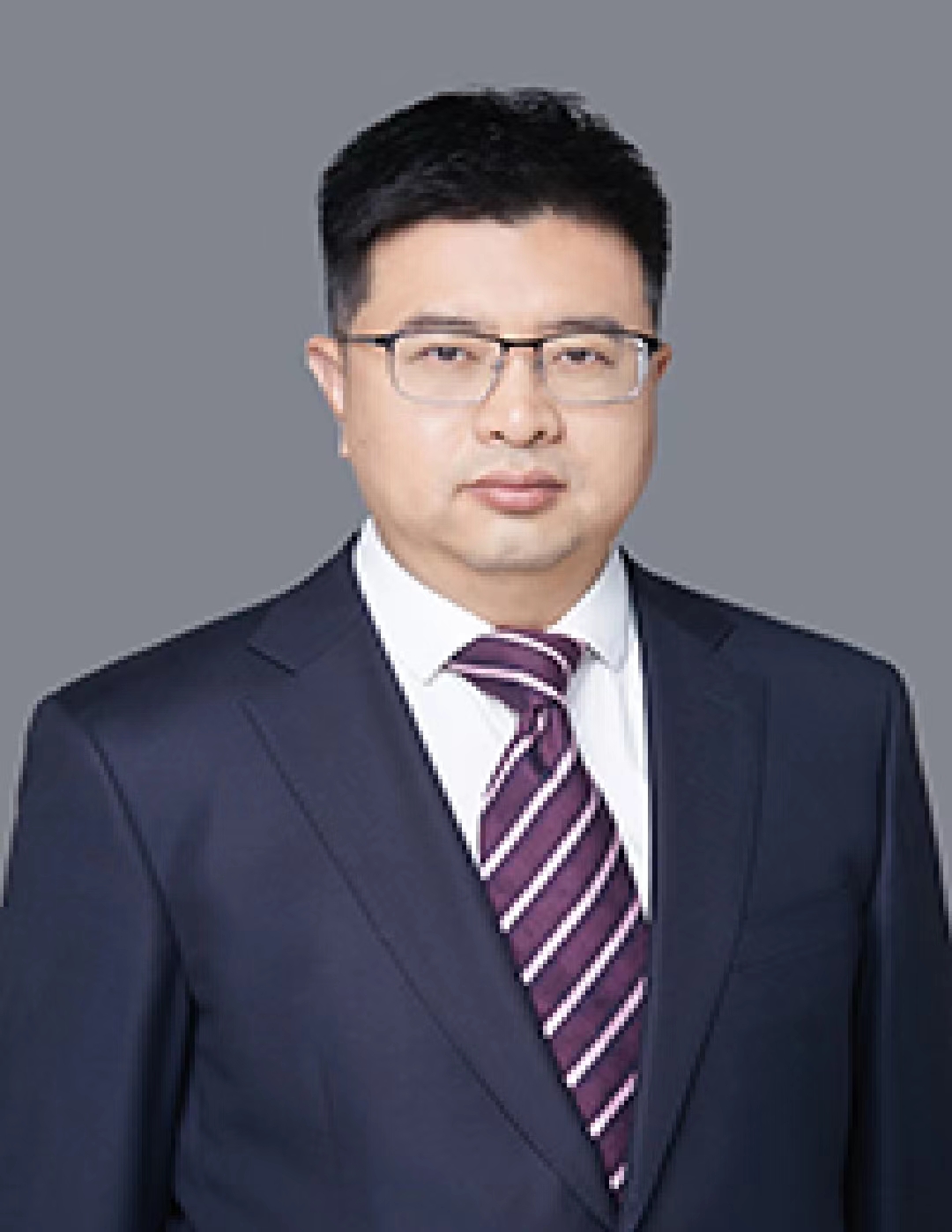}}]{Liehuang Zhu}
	received his Ph.D. degree in computer science from Beijing Institute of Technology, Beijing, China, in 2004. He is currently a professor at the School of Cyberspace Science and Technology, Beijing Institute of Technology. His research interests include security protocol  analysis and design, group key exchange protocols, wireless sensor networks, cloud computing, and blockchain applications.
\end{IEEEbiography}

%

\end{document}